\documentclass[12pt]{article}

\usepackage{epsfig}
\usepackage{graphicx}
\usepackage{amssymb,amsmath}
\usepackage{hyperref}
\usepackage[nosort]{cite}
\hypersetup{
    colorlinks=true,
    linkcolor=blue,
    citecolor=red,
}

\renewcommand{\Re}{\text{Re}\,}
\begin{document}

\begin{center}
\Large{\bf Lorentzian Vacuum Transitions with a Generalized Uncertainty Principle} \vspace{0.5cm}

\large  H. Garc\'{\i}a-Compe\'an\footnote{e-mail address: {\tt
compean@fis.cinvestav.mx}}, D. Mata-Pacheco\footnote{e-mail
address: {\tt dmata@fis.cinvestav.mx}}

\vspace{0.3cm}

{\small \em Departamento de F\'{\i}sica, Centro de
Investigaci\'on y de Estudios Avanzados del IPN}\\
{\small\em P.O. Box 14-740, CP. 07000, Ciudad de M\'exico, M\'exico}\\

\vspace*{1.5cm}
\end{center}

\begin{abstract}
The vacuum transition probabilities between to minima of a scalar field potential in the presence of gravity are studied using the Wentzel-Kramers-Brillouin approximation. First we propose a method to compute these transition probabilities by solving the Wheeler-DeWitt equation in a semi-classical approach for any model of superspace that contains terms of squared as well as linear momenta in the Hamiltonian constraint generalizing in this way previous results. Then we apply this method to compute the transition probabilities for a Friedmann-Lemaitre-Robertson-Walker metric with positive and null curvature and for the Bianchi III metric when the coordinates of minisuperspace obey a Standard Uncertainty Principle and when a Generalized Uncertainty Principle is taken into account. In all cases we compare the results and found that the effect of considering a Generalized Uncertainty Principle is that the probability is enhanced at first but it decays faster so when the corresponding scale factor is big enough the probability is reduced. We also consider the effect of anisotropy and compare the result of the Bianchi III metric with the flat FLRW metric which corresponds to its isotropy limit and comment the differences with previous works.

\vskip 1truecm

\end{abstract}

\bigskip

\newpage

\section{Introduction}
\label{S-Intro}
The search of a suitable quantum theory of gravity has been one of the main focus of theoretical physics for many decades\cite{Oriti:2009,Armas:2021}. There are some scenarios with the appropriate conditions where quantum gravity  is important to understand and correctly describe the physics involved. Much of these scenarios are encountered naturally in the context of the physics of the very early universe or the regions where it is expected that the curvature of spacetime increases as to create a singularity such as in the interior of a black hole or even the expected singularity at the big bang. One other scenario where this kind of theory is needed is the process of vacuum transitions in the presence of gravity since it is quantum in nature. These transitions were first studied using the path integral method in an Euclidean approach considering that each vacuum is defined by the minimum of a scalar field potential by Coleman and De Luccia \cite{ColemanDeLuccia} and later by Parke \cite{Parke:1982pm}. This process is described in field theory by the nucleation of true vacuum bubbles and the subsequent evolution is described after making an analytic continuation to the Lorentzian description and it is found that only open universes can arise as the end result. More recently, a different method to study these transitions was developed by using a Hamiltonian approach by Fishler, Morgan and Polchinski \cite{FMP1,FMP2} and it was used to study the transitions between two de Sitter universes and Minkowski to de Sitter in \cite{deAlwis:2019dkc}, it was found possible to describe transitions between closed universes contrary to the results using the Euclidean approach. It was conjectured that the Euclidean result could be only a limitation of relying on an analytic continuation. 

Using a Hamiltonian approach and canonical quantization to General Relativity is one of the earliest proposals to pursue a quantum theory of gravity \cite{Arnowitt:2008}. The classical Hamiltonian  constraint is canonically quantized and results in the so called Wheeler-DeWitt (WDW) equation \cite{Wheeler,DeWitt:1967}. The solutions of this equation are wave functionals defined in superspace called wave functions of the universe in quantum cosmology (for some reviews, see \cite{Halliwell:2009,VargasMoniz:2010zz,Bojowald:2020nwa,review2021}). Although the interpretation of these solutions is still an ongoing debate (see for example \cite{Valentini:2021izg} for a recent discussion) the ratio of two solutions can be interpreted as a probability to obtain a transition between both configurations. This is the interpretation used in the Hamiltonian approach cited earlier. In these works there were considered only transitions between universes with a different cosmological constant. A generalization to consider explicitly a scalar field with the vacuum states described by the minima of its potential was carried out in \cite{Cespedes:2020xpn}. In order to solve the WDW equation a Wentzel-Kramers-Brillouin (WKB) approximation was used and the transition probability of a closed FLRW metric was computed explicitly. The results found were compatible with the ones obtained by using an Euclidean approach. However, since the minisuperspace approximation was used in this approach there is no description of bubble nucleation and therefore it is conjectured that this approach could describe a generalization of the tunnelling from nothing scenario. Later a generalization of these results to include anisotropic metrics was done in \cite{LVTAU} and also to include the Ho\v{r}ava-Lifshitz theory of gravity in \cite{LVTHL}. A proposal to leave the minisuperspace limitation was developed in \cite{Oshita:2021aux}.

In the context of quantum gravity it is common that some typical behaviours arise such as the problem of time which appears in the WDW equation as the lack of a time variable. Another behaviour that is expected in this context is the appearance of a minimal measurable length since it is expected that the standard classical view of spacetime as a smooth manifold breaks down and a more fundamental structure appears. In this regard, it is possible to obtain this behaviour by modifying the usual Heisenberg uncertainty principle leading to a Generalized Uncertainty Principle (GUP). This modification to the theory of General Relativity has been studied in great detail in many scenarios focusing on its classical as well as quantum behaviours \cite{Kempf:1994su,Anacleto:2015mma,Anacleto:2015rlz,Scardigli:2016ubl,Scardigli:2016pjs,Lambiase:2017adh,Vagenas:2017fwa,Bosso:2017hoq,Demir:2018akw,Bosso:2018ckz,Fu:2021zrd,Bushev:2019zvw,Ali:2015ola,Pramanik:2014mma,Faizal:2015Dir,Faizal:2016zlo,Masood:2016wma}. Since this is an expected quantum feature it is natural to consider a GUP in superspace and look for solutions to the WDW equation. This has been studied in different scenarios before \cite{Vakili:2007yz,Vakili:2008tt,Kober:2011uj,Zeynali:2012tw,Faizal:2014rha,Faizal:2015kqa,Garattini:2015aca,Gusson:2020pgh,GUP,GUPHL}. In particular in \cite{GUP,GUPHL} it was studied the effects on General Relativity as well as the 
Ho\v{r}ava-Lifshitz theory for a Kantowski-Sachs metric. It was found that the modification of the WDW equation is not trivial but a WDW equation that takes into consideration the GUP can be found starting from the theory with standard uncertainty principle (SUP). The incorporation of a GUP into the WDW equation lets us incorporate a quantum behaviour which is interesting to study. In particular, in the present article we will examine the effect of the GUP in the vacuum transition probabilities when we take the coordinates on minisuperspace to be the gravitational degrees of freedom and the scalar field. 

When one is considering a quantum cosmological set-up it is common to use an FLRW metric since it fulfils  the cosmological principle and it is the easiest metric to explore in the picture of the WDW equation since it only has one gravitational degree of freedom, namely the scale factor. However, recent observational studies challenges the general idea that our universe possess a perfect level of isotropy leading to a possible degree of anisotropy  \cite{Colin:2018ghy,Migkas:2020fza,Migkas:2021zdo,Krishnan:2021dyb,Krishnan:2021jmh}. Therefore there is still an ongoing debate in this matter and thus considering metrics that are homogeneous but anisotropic is important and well justified with experimental data in the context of quantum cosmology as well. Therefore in the present work we are going to consider both kind of metrics, the FLRW metric and the anisotropic Bianchi III metric so we do not only study the effect of considering a GUP in the transition probability with the FLRW metric but also the effect of considering a GUP and anisotropy.

The organization of this article is as follows. In Section \ref{S-GeneralMethod} we present the general method to compute the transition probabilities when the WDW equation contain terms of linear momenta that generalizes the method presented in \cite{Cespedes:2020xpn,LVTAU}. The method is presented in a general form but it will be useful in this work for the specific case when we consider a GUP. Then we proceed to apply this method for the cases of an FLRW metric with positive curvature in Section \ref{S-FLRWClosed}, an FLRW metric with zero curvature in Section \ref{S-FLRW} and a Bianchi III metric in Section \ref{S-Bianchi}. In all cases we compute the transition probabilities when the coordinates obey a SUP as well as when a GUP is taken into account and compare the results. We also compare the Bianchi III metric result with the one obtained with the FLRW flat metric which represents its isotropy limit and comment on the comparison with the results found in \cite{LVTAU}. In Section \ref{GUP-EUP} we briefly discuss other extensions of the uncertainty principle that could lead to interesting results as well as the possible applicability of the general method. Finally in Section \ref{S-FinalRemarks} we give our final remarks.

\section{Semi-classical vacuum transition probabilities}
\label{S-GeneralMethod}
In this section we introduce a general method to compute the transition probabilities between two minima of a scalar field potential by obtaining a solution to a general WDW equation that contains quadratic as well as linear momentum terms using a WKB semi-classical approximation. This approach was developed in \cite{Cespedes:2020xpn,LVTAU} for a Hamiltonian with only terms of a quadratic dependence on momenta and therefore the method introduced in the present section represents its generalization. 

Let us consider a general Hamiltonian constraint written in the context of the ADM formalism of general relativity \cite{Arnowitt:2008,Wheeler,DeWitt:1967,Halliwell:2009,VargasMoniz:2010zz,Bojowald:2020nwa} of the form
\begin{equation}\label{GeneralHamiltonian}
		H=\frac{1}{2}G^{MN}(\Phi)\pi_{M}\pi_{N}+W^{M}(\Phi)\pi_{M}+f(\Phi)\simeq 0 ,
\end{equation}
where the degrees of freedom of the metric and the matter field variables are denoted by $\Phi^M$ (collectively they are denoted just by $\Phi$), and their canonical conjugate momenta are $\pi_{M}$. They define the space of variables with inverse metric $G^{MN}(\Phi)$ which will be the superspace of Wheeler after quantization. In this space a vector is defined with entries $W^{M}(\Phi)$ that depend in general on all the fields. Lastly $f(\Phi)$ is a scalar function that will be constructed by the scalar field potential, and all other terms that could arise that does not have a dependence on the momenta. We proceed to quantize the Hamiltonian by doing a standard canonical quantization procedure, that is substituting $\pi_{M}\to-i\hbar\frac{\delta}{\delta\Phi^M}$ in (\ref{GeneralHamiltonian}) from where we obtain
	\begin{equation}\label{WDWEqGeneral}
		H\Psi(\Phi)=\left[-\frac{\hbar^2}{2}G^{MN}(\Phi)\frac{\delta}{\delta\Phi^M}\frac{\delta}{\delta\Phi^N}-i\hbar W^M(\Phi)\frac{\delta}{\delta\Phi^M}+f(\Phi)\right]\Psi(\Phi)\simeq0 ,
	\end{equation}
where $\Psi(\Phi)$ is the wave functional defined on superspace and the resulting equation is called the WDW equation. In the quantization procedure some ambiguities related to the factor ordering can arise, however since we are going to use a semi-classical approximation these ambiguities will not be of importance. In order to have a semi-classical approximation to the solutions of the WDW equation we are going to propose an ansatz of the WKB type, that is we propose
	\begin{equation}\label{WKBAnsatz}
		\Psi(\Phi)=\exp\left\{\frac{i}{\hbar}S[\Phi]\right\} ,
	\end{equation}
where $S=S[\Phi]$ is a function that can be expanded as is usual in the WKB ansatz
	\begin{equation}\label{ActionExpansionh}
		 S[\Phi]=S_{0}[\Phi]+\hbar S_{1}[\Phi]+\mathcal{O}({\hbar}^2) ,
	\end{equation}
where $S_{0}$ is the classical action and $S_{1}$ will correspond to the first quantum correction term. In the context of the WDW equation taking a semi-classical approximation is similar to an s-wave approximation where we expect to capture some quantum aspects of the theory and still be able to relate them to the classical behaviour. We will see that this approximation is sufficient to study the vacuum transitions probabilities. Substituting this ansatz back in Eq. (\ref{WDWEqGeneral}) we obtain for the first two orders in $\hbar$ the following set of equations
	\begin{equation}\label{WDWFirstH}
		\frac{G^{MN}(\Phi)}{2}\frac{\delta S_{0}}{\delta\Phi^M}\frac{\delta S_{0}}{\delta\Phi^N}+W^M(\Phi)\frac{\delta S_{0}}{\delta\Phi^M}+f(\Phi)\simeq0 ,
	\end{equation}
	\begin{equation}\label{WDWSecondH}
		-\frac{i}{2}G^{MN}(\Phi)\frac{\delta^2S_{0}}{\delta\Phi^M\delta\Phi^N}+G^{MN}(\Phi)\frac{\delta S_{0}}{\delta\Phi^M}\frac{\delta S_{1}}{\delta\Phi^N}+W^{M}(\Phi)\frac{\delta S_{1}}{\delta\Phi^M}\simeq 0 .
	\end{equation}
We will focus only at the semi-classical level, that is only up to first order in $\hbar$ described by Eq. (\ref{WDWFirstH}). If we choose a different ordering in (\ref{WDWEqGeneral}) the modifications in the WDW equation will appear at second order in $\hbar$ and therefore at this level of approximation we can ignore any related issues. As it was carried out in \cite{Cespedes:2020xpn,LVTAU} we define a set of integral curves parametrized by a parameter $s$ and defined by
	\begin{equation}\label{IntegralCurves}
		C(s)\frac{d\Phi^M}{ds}=G^{MN}(\Phi)\frac{\delta S_{0}}{\delta\Phi^N} ,
	\end{equation}
where $C(s)$ is a function to be determined. We have the freedom to incorporate this function $C(s)$ in Eq. (\ref{IntegralCurves}) and it will allow us to take into account the information coming from the WDW equation. Since from (\ref{WDWFirstH}) we can obtain
	\begin{equation}\label{System1}
		G_{MN}(\Phi)\frac{d\Phi^M}{ds}\frac{d\Phi^N}{ds}=-\frac{2}{C(s)}\left[\frac{f(\Phi)}{C(s)}+W_{P}(\Phi)\frac{d\Phi^P}{ds}\right] ,
	\end{equation}
where $G_{MN}$ is the inverse of $G^{MN}$ and we lower and rise indices with this metric, that is $W_{P}=G_{LP}W^{L}$. Using (\ref{WDWFirstH}) we can also write down the classical action in the form
	\begin{equation}\label{ClassicalAction}
		S_{0}=-2\int_{X}\int_{s}ds\left[\frac{f(\Phi)}{C(s)}+W_{N}(\Phi)\frac{d\Phi^N}{ds}\right] ,
	\end{equation}
where $X$ is the spatial slice being considered. We note that Eqs. (\ref{IntegralCurves}) and (\ref{System1}) represent a system of equations for the variables $\frac{d\Phi^M}{ds}$ and $C(s)$. Therefore since we have the same number of equations as variables,  we have in principle all the information we need to solve this system and then substitute the results back into (\ref{ClassicalAction}) to obtain the classical action and with this the form of the wave functional solution (\ref{WKBAnsatz}) to the WDW equation up to first order in $\hbar$.

Let us assume that all the fields on superspace depend only on the time variable. Then we can compute the variational derivative of the classical action, in this way (\ref{IntegralCurves}) takes the form
	\begin{multline}\label{System2}
		\frac{d\Phi^M}{ds}=-\frac{2{\rm Vol}(X)}{C^2(s)}G^{MN}\frac{\partial f}{\partial\Phi^N}-\frac{2{\rm Vol}(X)}{C(s)}\left[\left(G^{MN}G_{LP}\frac{\partial W^{P}}{\partial\Phi^N}-\frac{\partial W^M}{\partial\Phi^L}\right)\frac{d\Phi^L}{ds}\right. \\ \left. +G^{MN}W^{P}\left(\frac{\partial G_{LP}}{\partial\Phi^N}-\frac{\partial G_{NP}}{\partial\Phi^L}\right)\frac{d\Phi^L}{ds}\right] ,
	\end{multline}
where  Vol$(X)$ represents the volume of the spatial slice. The system of equations composed by (\ref{System1}) and (\ref{System2}) reduces to the system considered in \cite{LVTAU} when $W^{M}=0$, where a general solution exists and the fields on superspace were found to be not independent to each other. In this case however a general solution is more difficult to be found and we will only solve the system in the special cases that will be considered in the following sections.

The interpretation of the solutions of the WDW equation is still an ongoing debate. One of the possible interpretations proposed is that the squared ratio of two solutions represent a relative probability of obtaining a certain universe configuration with respect to the other. With this framework in mind in \cite{Cespedes:2020xpn} the transition probability between two minima of the scalar field potential was proposed to be obtained by
	\begin{equation}\label{DefProb}
		P(A\to B)=\bigg|\frac{\Psi(\varphi^I_{0},\phi_{B};\varphi^I_{f},\phi_{A})}{\Psi(\varphi^I_{0},\phi_{A};\varphi^I_{f},\phi_{A})}\bigg|^2,
	\end{equation}
where $\phi_{A}$ and $\phi_{B}$ are the values that the scalar field takes in each minimum. In the following we will consider that $\phi_{A}$ represents the false minimum and $\phi_{B}$ the true one and then consider transitions from a false minimum to a true one. In the above expression $\Psi(\varphi^I_{0},\phi_{B};\varphi^M_{f},\phi_{A})$ represents the path in field space in which the scalar field changes from $\phi_{B}$ to $\phi_{A}$ and all the others fields in superspace evolve from an initial value $\varphi^I_{0}$ to a final value  $\varphi^I_{f}$ (we have separated here the scalar field $\phi$ from all the other fields in superspace denoted by $\varphi^{I}$). In the other hand $\Psi(\varphi^I_{0},\phi_{A};\varphi^M_{f},\phi_{A})$ represents the path in which the scalar field remains constant whereas all the other fields evolve. We will consider that this path is specified by the parameter $s$ used in (\ref{IntegralCurves}) taking values in the interval $[0,s_{f}]$. Thus taking the WKB ansatz proposed previously we obtain that the transition probability will be expressed as
	\begin{equation}\label{Probability}
		P(A\to B)= \exp(-2\Re[\Gamma]) ,
	\end{equation}
where 
	\begin{equation}\label{DefGamma}
		\pm\Gamma=\frac{i}{\hbar}\left[S_{0}(\varphi^I_{0},\phi_{B};\varphi^I_{f},\phi_{A})-S_{0}(\varphi^I_{0},\phi_{A};\varphi^I_{f},\phi_{A})\right] ,
	\end{equation}
the sign ambiguity comes from the fact that the general solution in (\ref{DefProb}) will consist of a linear superposition of exponentials. However we can only keep the dominant term. In this form, the transition probability is written in the same way as in \cite{Cespedes:2020xpn,LVTAU} but the system of equations and the expression of the classical action are more complicated due to the presence of the vector $W^{M}$.

It is worth mentioning that we are working in General Relativity (GR) and we will consider a GUP in the following sections. GR is an effective theory that is badly behaved in the ultraviolet (high energies) thus it has to be substituted by an appropriate ultraviolet completion. However GR is a very good approach to which quantum theories of gravity has to be reduced in the low energy limit.  It is possible to follow a conservative bottom-top approach by starting from GR and consider the lowest quantum effects ($s$-wave) as the quantum cosmology in the minisuperspace and then implement a GUP. In this case the semiclassical approximation would still capture some features of the microscopic spacetime. In the present case we are using this approach following Refs. \cite{Cespedes:2020xpn,LVTAU}, where it was studied the WKB semi-classical approximation which is known to give the leading contributions by instantons to the tunnelling between two minima of the double well potential using Euclidean path integrals in quantum mechanics and quantum field theory \cite{ColemanDeLuccia}. Higher order corrections in $\hbar$ in the WKB approximation would then describe quantum corrections to the saddle point. This is the general view that we pursue in the present paper in the computation of transition probabilities.

\section{Transitions in a closed FLRW metric}
\label{S-FLRWClosed}
We will begin by computing the transition probability in an homogeneous and isotropic universe represented with the FLRW metric with positive curvature when the variables on minisuperpace obey the usual commutation relations. The metric is written in the form
	\begin{equation}\label{FLRWMetric}
			ds^2=-N^2(t)dt^2+a^2(t)\left(dr^2+\sin^2\theta d\Omega^2_{2}\right) ,
	\end{equation}
where $N(t)$ is the lapse function, $d\Omega^2_{2}$ is the metric of
a $2$-sphere and $0\leq r\leq \pi$. Using natural units in which $c=1$ and $8\pi G$=1 and from now on $\hbar=1$\footnote{These units will be used throughout this work.} and considering gravity with a canonically coupled scalar field with potential $V(\phi)$ we obtain  the Hamiltonian constraint 
	\begin{equation}\label{FLREHamAux}
		H=N\left[\frac{\pi^2_{\phi}}{2a^3}-\frac{\pi^2_{a}}{12a}-3a+a^3V(\phi)\right]\approx
	0 ,
	\end{equation}
where the volume of the spatial slice is in this case Vol$(X)=2\pi^2$ which is only a global multiplicative factor. This Hamiltonian was studied in \cite{LVTAU} and the transition probabilities were obtained by using the general method  considered here but with $W^{M}=0$. However for this article it will be more useful to write everything in terms of the new variable $A(t)$ defined by $a(t)=e^{A(t)}$. Therefore the Hamiltonian constraint takes the form
	\begin{equation}\label{FLRWHC}
		H=Ne^{-3A}\left[-\frac{\pi^2_{A}}{12}+\frac{\pi^2_{\phi}}{2}+e^{6A}V(\phi)-3e^{4A}\right]\simeq0 ,
	\end{equation}
where the canonical momenta are
	\begin{equation}\label{FLRWMomenta}
		\pi_{\phi}=\frac{e^{3A}}{N}\dot{\phi} , \hspace{1cm} \pi_{A}=-\frac{6e^{3A}}{N}\dot{A} .
	\end{equation}
Since the lapse function is only a Lagrange multiplier, we can focus only on the term inside brackets in expression (\ref{FLRWHC}). Furthermore, since we factorized a global factor of $e^{-3A}$ we expect the transition probability to be slightly different that the one obtained in \cite{Cespedes:2020xpn,LVTAU} because both expressions are equivalent at the classical level but at the quantum level they may differ. We note that this Hamiltonian has the general form (\ref{GeneralHamiltonian}) with the fields in superspace $\{a,\phi\}$, inverse metric diagonal with $G^{AA}=-\frac{1}{6}$, $G^{\phi\phi}=1$, $f(A,\phi)=e^{6A}V(\phi)-3e^{4A}$ and $W^{M}=0$. Therefore we can follow the general solution presented in \cite{LVTAU} to compute the transition probability. Thus we propose that the parameter $s$ is defined such that
	\begin{equation}\label{ChooseS}
		\phi(s) \approx
		\begin{cases}
			\phi_{B} , & 0<s<\bar{s}-\delta s,\\
			\phi_{A} , & \bar{s}+\delta s<s<s_{M} ,
		\end{cases}
	\end{equation}
and the thin wall limit will be represented by $\delta s\to0$. With this definition the actions involved in the transition can be written as
	\begin{multline}
		S_{0}(a_{0},\phi_{B};a_{f},\phi_{A})=-4\pi^2\left[\int_{0}^{\bar{s}-\delta s}\frac{f}{C(s)}ds\bigg\rvert_{\phi=\phi_{B}}+\int_{\bar{s}-\delta s}^{\bar{s}+\delta s}\frac{f}{C(s)}ds\right. \\ \left.+\int_{\bar{s}+\delta s}^{s_{M}}\frac{f}{C(s)}ds\bigg\rvert_{\phi=\phi_{A}}\right] ,
	\end{multline}
	\begin{equation}
		S_{0}(a_{0},\phi_{A};a_{f},\phi_{A})=-4\pi^2\int_{0}^{s_{M}}\frac{f}{C(s)}ds\bigg\rvert_{\phi=\phi_{A}} .
	\end{equation}
Then by using (\ref{DefGamma}) we obtain
	\begin{multline}\label{FLRWGammaA}
		\pm\Gamma=-4\pi^2i\left[\int_{0}^{\bar{s}-\delta s} \frac{f}{C(s)}\bigg\rvert_{\phi=\phi_{B}}ds-\int_{0}^{\bar{s}-\delta s}\frac{f}{C(s)}\bigg\rvert_{\phi=\phi_{A}}ds \right] \\-4\pi^2i\int_{\bar{s}-\delta s}^{\bar{s}+\delta s}\frac{ds}{C(s)}e^{6A}\left[V(\phi)-V_{A}\right].
	\end{multline}
The two first integrals can be evaluated by solving the system of equations in the case in which the scalar field is constant and making a change of variables it is possible to carry out the integrals on the scale factor. On the other hand the last term can be written as a tension term in the form
	\begin{equation}\label{FLRWTensionDef}
		2\pi^2e^{6\bar{A}}T=-4\pi^2i\int_{\bar{s}-\delta s}^{\bar{s}+\delta s}\frac{ds}{C(s)}e^{6A}\left[V(\phi)-V_{A}\right] ,
	\end{equation}
where $\bar{A}=A(s=\bar{s})$. Therefore we finally obtain in the thin wall limit
	\begin{equation}\label{FLRWTRP}
		\pm\Gamma=\pm12\pi^2\bigg\{\frac{1}{V_{B}}\bigg[\left(1-\frac{V_{B}}{3}e^{2\bar{A}}\right)^{3/2}-1\bigg]-\frac{1}{V_{A}}\bigg[\left(1-\frac{V_{A}}{3}e^{2\bar{A}}\right)^{3/2}-1\bigg]\bigg\}+2\pi^2e^{6\bar{A}}T ,
	\end{equation}
where the sign ambiguity on the right hand side comes from the fact that the general solution of the system of equations gives a solution for $C^2(s)$ and therefore it is independent to the ambiguity in the left hand side. We have also performed the integral by taking the initial value to be $A_{0}=A(s=0)=-\infty$ since we found no divergences there. We note that this expression differs from the results found in \cite{Cespedes:2020xpn,LVTAU} only on the power of the term accompanying the tension. This is because of the global factor factorized in (\ref{FLRWHC}) and represents a difference similar to what may arrive  by using different ordering. 

The transition probability found is described by two parameters $T$ and $\bar{A}$. However, we can relate them by looking from an extremum of this expression with respect to $\bar{A}$. Carrying out this procedure we obtain that the tension can be written as
	\begin{equation}\label{FLRWTension}
		T=\mp e^{-4\bar{A}}\left[\sqrt{1-\frac{V_{A}}{3}e^{2\bar{A}}}-\sqrt{1-\frac{V_{B}}{3}e^{2\bar{A}}}\right] ,
	\end{equation}
which looks like the value found in \cite{LVTHL} with an extra term of $e^{-3\bar{A}}/2$ because of the term that was factorized as explained before. As it was explained in that work, the tension term will only be well defined if the terms inside the square roots are positive, which can be fulfilled if the potentials are negative. If the potentials are positive they create an upper bound for the scale factor and thus to the validity of this expression. Then we obtain finally the transition probability described in terms of just one parameter as follows
	\begin{multline}\label{FLRWTRPF}
		\pm\Gamma=\pm12\pi^2\left\{\frac{1}{V_{B}}\left[\left(1-\frac{V_{B}}{3}e^{2\bar{A}}\right)^{3/2}-1\right]-\frac{1}{V_{A}}\left[\left(1-\frac{V_{A}}{3}e^{2\bar{A}}\right)^{3/2}-1\right]\right. \\ \left. -\frac{e^{2\bar{A}}}{6}\left[\sqrt{1-\frac{V_{A}}{3}}e^{2\bar{A}}-\sqrt{1-\frac{V_{B}}{3}}e^{2\bar{A}}\right]\right\} .
	\end{multline}

This expression differs from the result found in \cite{LVTAU,LVTHL} only by a constant factor regarding the last term. As it was explained in that work, the correct choices of sign in order to obtain the same result as the one obtain by using Euclidean methods is the plus sign in the right and in order to have a well defined probability a plus sign in the left. However since in this case the second term is smaller, it is not always possible to obtain a positive quantity in the right hand side if we choose the plus sign, therefore in this case in order to have a well defined probability we will choose the minus sign in the right hand side and the plus sign in the left.	

As it was mentioned in Section \ref{S-GeneralMethod} the ambiguity related to the factor ordering does not appear at the semi-classical level used in the present article. However as we will show next a term with linear momenta among other modifications will arise naturally when a GUP is taken into account and thus in order to compute the transition probabilities in this scenario the method of \cite{LVTAU} will not be sufficient. Instead, we must use the method introduced here in Section \ref{S-GeneralMethod}.

\subsection{Introducing the GUP}
Now that we have computed the transition probability when the coordinates in minisuperspace obey the standard uncertainty relations let us study the changes made when considering a GUP. We will consider  a commutation relation that leads to a non-zero minimum uncertainty in the position variable inspired by \cite{Kempf:1994su}, and used in \cite{GUP,GUPHL} in the context of the WDW equation. That is we will consider a set of coordinates $\{\Phi^{M}_{g}\}$ that obey
	\begin{equation}\label{RelationGUP}
		[\Phi^{M}_{g},\pi_{N}]=i\delta_{MN}\left(1+\gamma^2\mathcal{P}^2\right),
	\end{equation}
where $\gamma$ is a small parameter with units of
inverse momentum and $\mathcal{P}^2$ is the part of the WDW equation that contains all the quadratic terms in momenta, in this case we have
	\begin{equation}\label{FLRWP2}
		\mathcal{P}^2=-\frac{\pi^2_{A}}{12}+\frac{\pi^2_{\phi}}{2} .
	\end{equation}	
We will also define a set of coordinates in minisuperspace $\Phi^{M}$ that obey the usual commutation relations, that is $[\Phi^M,\pi_{N}]=i\delta_{MN}$. This set of coordinates corresponds to the one used so far. In position space both sets of coordinates can be related as
\begin{equation}\label{RelationsCoordinates}
	\Phi^{M}_{g}=\Phi^M\left(1+\gamma^2\mathcal{P}^2\right).
\end{equation}
We can also choose the following representation for the momentum operators
\begin{equation}\label{MomOperatorsGUP}
	\pi_{A}=-i\frac{\delta}{\delta A} , \hspace{1cm} \pi_{\phi}=-i \frac{\delta}{\delta\phi}.
\end{equation} 

As explained in \cite{GUP,GUPHL} in order to obtain a WDW equation with this set up we start by considering that Eq. (\ref{FLRWHC}) is valid written in terms of coordinates with the $g$ subscript and then each term in that expression has to be rewritten in terms of the coordinates without subscript by using the relations (\ref{RelationsCoordinates}). In order to rewrite each exponential term we are going to use the Zassenhaus formula which is written as 
	\begin{equation}\label{ZSFormula}
		e^{A+B}=e^{A}e^{B}e^{-\frac{1}{2}[A,B]}e^{\frac{1}{6}\left([A,[A,B]]+2[B,[A,B]]\right)  + \cdots},		
	\end{equation}
where $\cdots$ denote terms with commutators involving more than 3
operators. Since $\gamma$ is a small parameter in all the computations we are only going to write up to second order in $\gamma$. We will also consider up to second order in momenta. Therefore we obtain
	\begin{equation}\label{FLRWGUOTRAux}
		-3e^{4A_{g}}\Psi(A_{g},\phi_{g})\simeq -3e^{4\left(1+\frac{4\gamma^2}{9}\right)A}e^{4\gamma^2A\mathcal{P}^2}e^{\frac{4\gamma^2}{3}iA\pi_{A}}\Psi\left(A,\phi\right) .
	\end{equation}	
As it was discussed in \cite{GUPHL} there are two ways to proceed. The first one is to note that the exponential with the linear term in momentum acts as a rescaling of the variable $A$ in the wave functional. On the other hand the exponential term can be expanded up to second order in $\gamma$ and leave the dependence on the wave functional with the coordinate unchanged. In both cases the exponential term with $\mathcal{P}^2$ is also expanded in a power series. In this article we will consider in all cases the second option. Thus we obtain
	\begin{equation}\label{FLRWExpansion1}
		-3e^{4A_{g}}\Psi(A_{g},\phi_{g})\simeq -3e^{4\left(1+\frac{4\gamma^2}{9}\right)A}\left(1+\frac{4}{3}i\gamma^2A\pi_{A}+4\gamma^2A\mathcal{P}^2\right)\Psi\left(A,\phi\right) 
	\end{equation}	
and in the same way using a Taylor expansion on the scalar field potential, we obtain up to second order in $\gamma$ 
	\begin{equation}\label{FLRWExpansion2}
		e^{6A_{g}}V(\phi_{g})\Psi(A_{g},\phi_{g})\simeq e^{6(1+\gamma^2)A}V(\phi)\left(1+3i\gamma^2A\pi_{A}+6\gamma^2A\mathcal{P}^2\right)\Psi\left(A,\phi\right) .
	\end{equation}
Therefore the resulting WDW equation takes the form
	\begin{multline}\label{FLRWGUPHC}
	\left\{\left[1+6\gamma^2e^{6(1+\gamma^2)A}V(\phi)A-12\gamma^2e^{4\left(1+\frac{4}{9}\gamma^2\right)A}A\right]\left(-\frac{\pi^2_{A}}{12}+\frac{\pi^2_{\phi}}{2}\right) \right. \\ \left. +\left[3i\gamma^2e^{6(1+\gamma^2)A}AV(\phi)-4i\gamma^2e^{4\left(1+\frac{4\gamma^2}{9}\right)A}A\right]\pi_{A}  \right. \\ \left. +e^{6(1+\gamma^2)A}V(\phi)-3e^{4\left(1+\frac{4\gamma^2}{9}\right)A}\right\}\Psi\left(A,\phi\right)\simeq0 .
	\end{multline}
We note that this expression is the same as the one considered in (\ref{GeneralHamiltonian}) with 
	\begin{equation}\label{FLRWGUPMetric}
		\{G^{MN}\}=\left(1+6\gamma^2e^{6(1+\gamma^2)A}V(\phi)A-12\gamma^2e^{4\left(1+\frac{4\gamma^2}{9}\right)A}A\right){\rm diag}\left(-\frac{1}{6},1\right), 
	\end{equation}
	\begin{equation}\label{FLRWGUPW}
		\{W^{M}\}=\left(3i\gamma^2e^{6(1+\gamma^2)A}AV(\phi)-4i\gamma^2e^{4\left(1+\frac{4\gamma^2}{9}\right)A}A,0\right),
	\end{equation}
	\begin{equation}\label{FLRWGUPf}
		f(A,\phi)=e^{6(1+\gamma^2)A}V(\phi)-3e^{4\left(1+\frac{4\gamma^2}{9}\right)A} .
	\end{equation}

The WDW equation obtained in this way has an extra term of linear momenta. This is not the only change, since all the other factors are also modified. We note that Eq. (\ref{FLRWGUPHC}) simplifies to the WDW equation obtained after quantizing (\ref{FLRWHC}) with standard ordering in the limit $\gamma\to 0$ and therefore this equation represents a non trivial deformation of the standard case depending on the $\gamma$ parameter which has the physical significance of measuring the deviation from the SUP. In contrast if we consider a different ordering in (\ref{FLRWHC}) we can also obtain a linear term depending on the factor ordering which has few physical meaning but the other terms will be unchanged and its effect will not be present at the level of approximation we are considering in the present paper. Thus taking into account the GUP is a more compelling scenario compared to the case of just considering a different ordering and on the contrary its effects will be present even at the semi-classical level used.

In order to study the transitions in this set up we choose the parameter $s$ as in (\ref{ChooseS}), then using (\ref{ClassicalAction}) and (\ref{DefGamma}) we obtain in this case
	\begin{multline}\label{FLRWGUPGamma}
		\pm\Gamma=-4\pi^2i\left[\int_{0}^{\bar{s}-\delta s} \left(\frac{f}{C(s)}+W_{M}\frac{d\Phi^M}{ds}\right)\bigg\rvert_{\phi=\phi_{B}}ds-\int_{0}^{\bar{s}-\delta s}\left(\frac{f}{C(s)}+W_{M}\frac{d\Phi^M}{ds}\right)\bigg\rvert_{\phi=\phi_{A}}ds \right] \\ +2\pi^2e^{6(1+\gamma^2)\bar{A}}T ,
	\end{multline}
where, as it was done in \cite{LVTAU}, we have defined the tension term as the portion of the integral where the scalar field can vary, that is
	\begin{equation}\label{FLRWGUPTensionDef}
		2\pi^2e^{6(1+\gamma^2)\bar{A}}T=-4\pi^2i\int_{\bar{s}-\delta s}^{\bar{s}+\delta s}\frac{ds}{C(s)}e^{6(1+\gamma^2)A}\left[V(\phi)-V_{A}\right] .
	\end{equation}
The two first integrals in (\ref{FLRWGUPGamma}) can be evaluated by solving the general system of equations (\ref{System1}) and (\ref{System2}) in the special case in which the scalar field is constant. Since in that case there is only one coordinate in minisuperspace the system simplifies and can be solved giving as solutions
	\begin{equation}\label{FLRWGUPAux}
		C(s)=\frac{2\pi^2}{f}\frac{\partial f}{\partial A}\left[W^{A}\pm\sqrt{(W^{A})^2-2fG^{AA}}\right] ,
	\end{equation}
	\begin{equation}\label{FLRWGUPAux2}
		\frac{dA}{ds}=-\frac{G^{AA}f^2}{\pi^2\frac{\partial f}{\partial A}\left[W^{A}\pm\sqrt{(W^{A})^2-2fG^{AA}}\right]^2} ,
	\end{equation}
where the sign ambiguity comes from the fact that the system leads to a quadratic equation for $C(s)$. In order to obtain all the results in terms of only $A$ we can perform a change of variables in the integral according to $dA=\left(\frac{dA}{ds}\right)ds$  which leads us to
	\begin{equation}\label{FLRWGUPAction}
		\int \left(\frac{f}{C(s)}+W_{M}\frac{d\Phi^M}{ds}\right)ds=\left[\frac{1}{2}\int\frac{W^{A}}{G^{AA}}dA\pm\int\frac{\sqrt{(W^{A})^2-2fG^{AA}}}{2G^{AA}}dA\right] .
	\end{equation}
  However we note that $W^{A}$ has an overall dependence on $\gamma^2$, therefore its squared value can be ignored comparing with $2fG^{AA}$, therefore the second integral simplifies and we finally obtain in the thin wall limit and up to second order in $\gamma$
	\begin{multline}\label{FLRWGUPProb}
		\pm\Gamma\simeq\pm12\pi^2\left[\int_{-\infty}^{\bar{A}}\sqrt{F(A,V_{B})}dA-\int_{-\infty}^{\bar{A}}\sqrt{F(A,V_{A})}dA\right] \\ -12\pi^2\gamma^2\left[\int_{-\infty}^{\bar{A}}G(A.V_{B}) dA-\int_{-\infty}^{\bar{A}}G(A,V_{A})dA\right] + 2\pi^2e^{6(1+\gamma^2)A}T ,
	\end{multline}
where we have defined the functions
	\begin{equation}\label{FLRWGUPDefF}
		F(A,V)=\frac{e^{4\left(1+\frac{4\gamma^2}{9}\right)A}-\frac{V}{3}e^{6(1+\gamma^2)A}}{1+6\gamma^2e^{6(1+\gamma^2)A}VA-12\gamma^2e^{4\left(1+\frac{4\gamma^2}{9}\right)A}A} ,
	\end{equation}
	\begin{equation}\label{FLRWGupDefG}
		G(A,V)=\frac{3Ve^{6(1+\gamma^2)A}-4e^{4\left( 1+\frac{4\gamma^2}{9}\right)A}}{1+6\gamma^2e^{6(1+\gamma^2)A}VA-12\gamma^2e^{4\left(1+\frac{4\gamma^2}{9}\right)A}A} A ,
	\end{equation}
and the limits of integration have been taken as in the case without a GUP. We note that for some values of the potentials the integrals of the $F$ function could result in imaginary contributions which will be ignored after taking the real part of $\Gamma$, however the  integral of the $G$ function does not have this problem and therefore they will always have to be taken into account. Thus the dependence on the $\gamma$ parameter will always be present. We note that this expression simplifies to (\ref{FLRWTRP}) in the limit $\gamma\to0$ as expected. The probability just found is still described by only two parameters, namely the tension and $\bar{A}$. As it was done in the latter case they can be related by an extremizing procedure, however we obtain in this case
	\begin{equation}\label{FLRWGUPTension}
		T=\frac{e^{-6(1+\gamma^2)\bar{A}}}{1+\gamma^2}\left[\gamma^2\left(G(\bar{A},V_{B})-G(\bar{A},V_{A})\right)\mp\left(\sqrt{F(\bar{A},V_{B})}-\sqrt{F(\bar{A},V_{A})}\right)\right] ,
	\end{equation}
and thus its validity will only be possible when $F(\bar{A},V_{A,B})\geq0$. We note however that when $\bar{A}>>1$ we have
	\begin{equation}
		F(A,V_{A,B})\to-\frac{1}{18\gamma^2\bar{A}} ,
	\end{equation}
and consequently this will represent an upper bound for $\bar{A}$ in which the expression (\ref{FLRWGUPTension}) could be well defined regardless of the value of the potential minima. Thus we are going to stay with the expression before this varying procedure (\ref{FLRWGUPProb}) and consider both parameters as independent. We also note that both functions (\ref{FLRWGUPDefF}), (\ref{FLRWGupDefG}) have a singularity in which the denominator goes to zero for some value of $A$. Therefore we are going to consider regions that do not contain this value so the integrals can be performed numerically.

\begin{figure}[h!]
	\centering
	\includegraphics[width=0.6\textwidth]{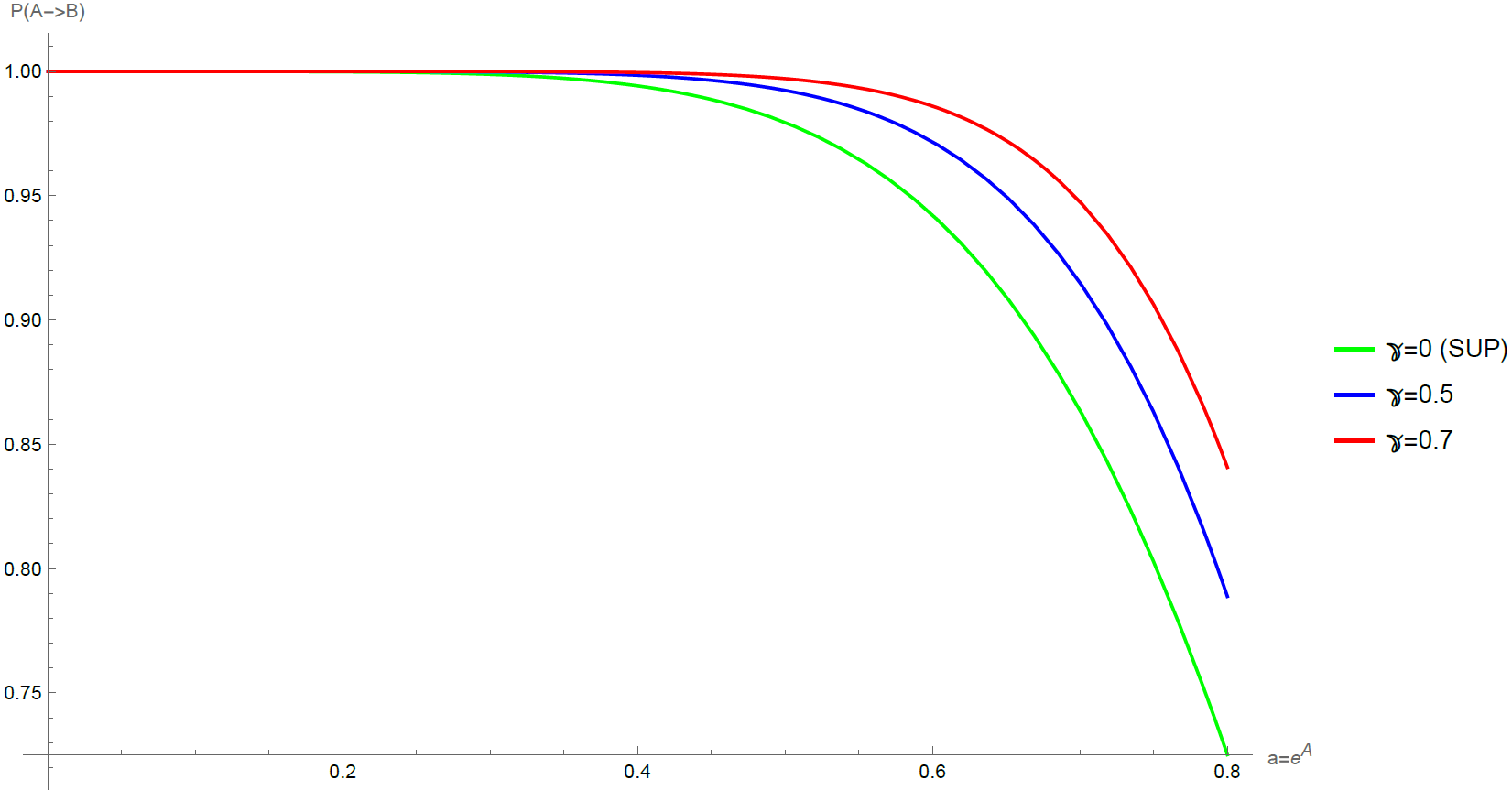}
	\caption{Transition probability for the FLRW closed metric in the UV region with SUP or $\gamma=0$ (Green curve), and considering a GUP with $\gamma=0.5$ (Blue curve) and $\gamma=0.7$ (Red curve) for $V_{A}=-1$, $V_{B}=-2$ and $T=7$.  }
	\label{F-FLRW}
\end{figure}

\begin{figure}[h!]
		\centering
		\includegraphics[width=0.6\textwidth]{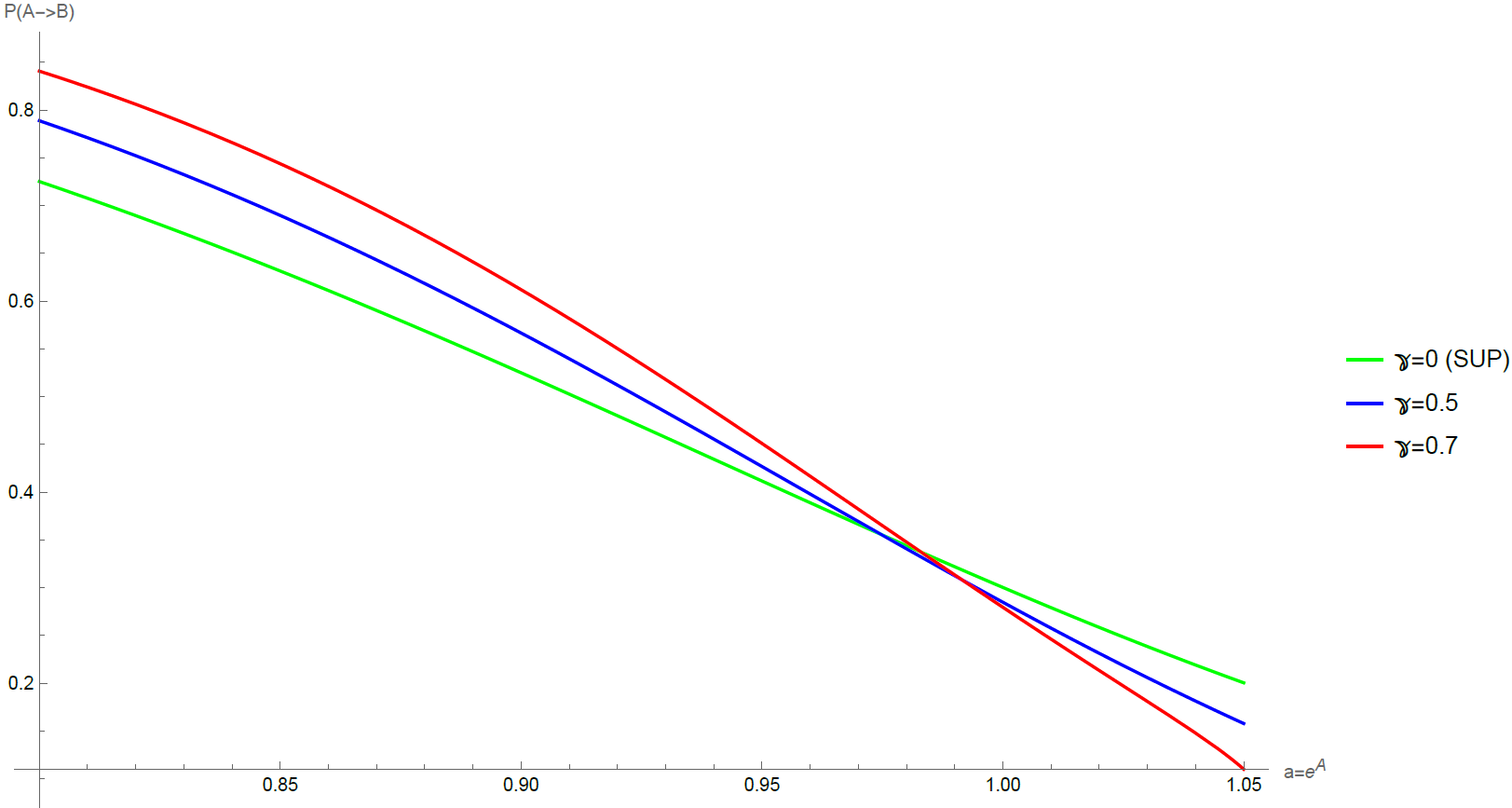}
		\caption{Transition probability for the FLRW closed metric with SUP or $\gamma=0$ (Green curve), and considering a GUP with $\gamma=0.5$ (Blue curve) and $\gamma=0.7$ (Red curve) for $V_{A}=-1$, $V_{B}=-2$ and $T=7$ focusing on the region where the change of behaviour occurs. }
		\label{F-FLRWExtra}
	\end{figure}

As it was explained before the only difference between the logarithm of the transition probability found here in (\ref{FLRWTRP}) and the result found in \cite{Cespedes:2020xpn,LVTAU} is the power of the scale factor in the tension term. In this case however for values of the scale factor near zero this term is much smaller than the term found when nothing is factorized. Thus since the first factor is negative if we choose the plus sign, this factor is bigger than the tension term in this region and that leads us to an ill defined probability. Therefore in this case we are going to choose the minus sign in the first factor and thus we can obtain a well defined probability. For the case when a GUP is considered (\ref{FLRWGUPProb}) we will also choose the sign that assure us to obtain well defined probabilities, in this case this is the plus sign (which corresponds to the minus sign in the previous case). With these considerations we can proceed to compare both results. We find that the general form of the probability is the same in all cases, that is they all start at $1$ in the limit $\bar{a}\to0$ and then go to $0$ as the values of the scale factor increases. However the impact of considering a GUP is that the probability in the UV region (small scale factor) is bigger but it decreases faster and thus in the IR region (big scale factor) the probability is smaller. Since the effect is subtle because $\gamma$ was taken to be a small number, in order to visualize in a great form the regions of interest we show two plots. In Figure \ref{F-FLRW} we show a plot of the UV region and in Figure \ref{F-FLRWExtra} we show a plot in the region where the change in the behaviour of the probabilities is appreciated. By doing this we show that the point where the change of behaviour is found is not the same for all the curves and thus there is a small intermediate region where the probability has neither an increasing nor decreasing behaviour with $\gamma$. In both cases we choose $V_{A}=-1$, $V_{B}=-2$, $T=7$, we ignore the factor $2(12\pi^2)$ since it is just a global factor in all the expressions and plot for different values of $\gamma$. In this case the region where a singularity could appear in the $F$ or $G$ function is in the right hand side of the figures, that is for bigger values of the scale factor where the probability is very small. 


\section{Transitions in a flat FLRW metric}
\label{S-FLRW}
Now that we have studied the transition probabilities in the case of a positive curvature let us study them with the FLRW metric without curvature. This metric has a noncompact spatial part, however it will be useful to compare with the result of the next section where we will investigate the effect of anisotropy.

The flat FLRW metric can be written in Cartesian coordinates as
	\begin{equation}\label{FLRWFMetric}
		ds^2=-N(t)dt^2+a^2(t)\left[dx^2+dy^2+dz^2\right] ,
	\end{equation}
and considering the coupling to a scalar field we obtain the Hamiltonian constraint
	\begin{equation}\label{FLRWFHCAux}
		H=\frac{N}{a^3}\left[\frac{\pi^2_{\phi}}{2}-\frac{a^2\pi^2_{a}}{12}+a^6V\right]\approx 0 .
	\end{equation}
In this article however it is more convenient to make the change of coordinate $a=e^{A(t)}$, and then the Hamiltonian constraint takes the form
	\begin{equation}\label{FLRWFHC}
		H=Ne^{-3A}\left[-\frac{\pi^2_{A}}{12}+\frac{\pi^2_{\phi}}{2}+e^{6A}V(\phi)\right]\simeq0 .
	\end{equation}
We can study the transition probabilities using the WDW equation obtained after quantizing the latter expression assuming standard commutation relations, for this we use the general procedure of \cite{LVTAU} choosing  $s$ as defined in (\ref{ChooseS}) as in the previous section we obtain
	\begin{equation}\label{FLRWFProb}	
		\pm\Gamma=\pm\frac{2{\rm Vol}(X)i}{\sqrt{3}}\left(\sqrt{V_{B}}-\sqrt{V_{A}}\right)e^{3\bar{A}}+{\rm Vol}(X)e^{6\bar{A}}T ,
	\end{equation}
where we have chosen $A_{0}=-\infty$ since it does not produce any divergence and used the thin wall limit. Since the spatial slice of the metric is not compact the term ${\rm Vol}(X)$ in the last expression is in principle a divergent number. However as explained in \cite{LVTAU} we could choose a correctly compactified spatial slice and then this term will have a finite value, in any case, the general form will not be altered and therefore in this result and in the following we will simply let this term as a global constant. We note that this expression is the same as the one obtained in \cite{LVTAU} after making the change of variable since in this case the factorization of the term $e^{-3A}$ in (\ref{FLRWFHC}) was also made in that work.

The transition probability (\ref{FLRWFProb}) is described in terms of two parameters. We could try to relate both of them as done in the previous section using an extremizing procedure, however if we vary that expression we obtain
	\begin{equation}\label{FLRWFTension}
		T=\mp\frac{\left(\sqrt{V_{B}}-\sqrt{V_{A}}\right)i}{\sqrt{3}e^{3\bar{A}}} ,		
	\end{equation}
which is only well defined if both potential minima are negative. Therefore in general we will consider that both parameters are independent and choose signs that allows us to obtain well defined probabilities.

\subsection{Considering a GUP}

Let us consider coordinates in minisuperspace that obey a GUP, that is we consider (\ref{RelationGUP}) and proceed as in the previous section. Using (\ref{FLRWExpansion2}) we obtain that the WDW equation in this case is rewritten as
	\begin{multline}\label{FLRWGUPFHC}
		\left\{\left[1+6\gamma^2e^{6(1+\gamma^2)A}V(\phi)A\right]\left(-\frac{\pi^2_{A}}{12}+\frac{\pi^2_{\phi}}{2}\right) +3i\gamma^2e^{6(1+\gamma^2)A}AV(\phi)\pi_{A}  \right. \\ \left. +e^{6(1+\gamma^2)A}V(\phi)\right\}\Psi\left(A,\phi\right)\simeq0 .
	\end{multline}
Comparing with the general form (\ref{GeneralHamiltonian}) we note that in this case 
	\begin{equation}\label{FLRWGUPFMetric}
		\{G^{MN}\}=\left(1+6\gamma^2e^{6(1+\gamma^2)A}V(\phi)A\right){\rm diag}\left(-\frac{1}{6},1\right), 
	\end{equation}
	\begin{equation}\label{FLRWGUPFW}
		\{W^{M}\}=\left(3i\gamma^2e^{6(1+\gamma^2)A}AV(\phi),0\right) , \hspace{0.5cm}
		f(A,\phi)=e^{6(1+\gamma^2)A}V(\phi) .
	\end{equation}
Therefore doing the same procedure and taking the parameter $s$ as in (\ref{ChooseS}) we obtain that the logarithm of the transition probability is written in the form 
	\begin{multline}\label{FLRWFGUPGamma}
		\pm\Gamma=-2i{\rm Vol}(X)\left[\int_{0}^{\bar{s}-\delta s} \left(\frac{f}{C(s)}+W_{M}\frac{d\Phi^M}{ds}\right)\bigg\rvert_{\phi=\phi_{B}}ds\right. \\ \left. -\int_{0}^{\bar{s}-\delta s}\left(\frac{f}{C(s)}+W_{M}\frac{d\Phi^M}{ds}\right)\bigg\rvert_{\phi=\phi_{A}}ds \right] +\frac{{\rm Vol}(X)}{\hbar}e^{6(1+\gamma^2)\bar{A}}T .
	\end{multline}
Solving the corresponding system of equations to compute the two remaining integrals we also obtain in this case that Eq.(\ref{FLRWGUPAction}) holds and here we also note that the term $(W^{A})^2$ can be ignored with respect to $2fG^{AA}$. Thus we finally obtain in the thin wall limit and up to second order in $\gamma$
	\begin{multline}\label{FLRWGUPFProb}
		\pm\Gamma\simeq\pm2\sqrt{3}i{\rm Vol}(X)\left[\int_{-\infty}^{\bar{A}}\sqrt{F_{f}(A,V_{B})}dA-\int_{-\infty}^{\bar{A}}\sqrt{F_{f}(A,V_{A})}dA\right] \\ -18{\rm Vol}(X)\gamma^2\left[\int_{-\infty}^{\bar{A}}G_{f}(A.V_{B}) dA-\int_{-\infty}^{\bar{A}}G_{f}(A,V_{A})dA\right] + {\rm Vol}(X)e^{6(1+\gamma^2)\bar{A}}T ,
	\end{multline}
where 
	\begin{equation}\label{FLRWGUPFDefFG}
		F_{f}(A,V)=\frac{Ve^{6(1+\gamma^2)A}}{1+6\gamma^2e^{6(1+\gamma^2)A}VA} , \hspace{0.5cm}
		G_{f}(A,V)=AF_{f}(A,V) .
	\end{equation}
We note that this transition probability is described again in terms of just two parameters that in general could be taken as independent since the extremizing procedure will lead to similar difficulties as before. We also note that the second term in (\ref{FLRWGUPFProb}) always contributes to the probability since it is always a real term whereas the first term could be ignored for example in the case in which both potential minima are positive since in that case the first term will give an imaginary contribution. Finally, we note that in the limit $\gamma\to0$ this expression reduces to Eq. (\ref{FLRWFProb})	as expected.

We note from (\ref{FLRWGUPFDefFG}) that if the potential minima are positive, then the first term in (\ref{FLRWGUPFProb}) will not contribute to the transition probability as it happens with the first term in (\ref{FLRWFProb}). However if the potential minima are negative, these terms will contribute and the $F_{f}$ and $G_{f}$ functions will have a singular point and thus in order to have well defined probabilities we have to consider regions that avoid these points. In order to obtain well defined probabilities we also choose the minus sign in the sign ambiguity of the right hand side and the plus sign on the left from Eq. (\ref{FLRWGUPFProb}). We find the same behaviour encountered earlier for the closed FLRW metric, that is, at first the bigger the value of $\gamma$, the bigger the probability but it decreases faster and therefore when the scale factor is big enough the probability decreases with $\gamma$. However the general behaviour for all the cases is the same, they all start at $1$ in the UV limit and falls to $0$.

\section{Transitions in a Bianchi III metric}
\label{S-Bianchi}
So far we have studied the transition probabilities only  with FLRW metrics that describe homogeneous and isotropic models of the universe and leads to a minisuperspace defined only by two coordinates, namely, the scale factor and the scalar field. However, the method presented in Section \ref{S-GeneralMethod} was introduced in a general way to any model in minisuperspace that have a Hamiltonian in the form (\ref{GeneralHamiltonian}) and whose fields depend only on the time variable. Therefore in this section we are going to study a more general metric that incorporates more fields in superspace. In particular we are going to use the Bianchi III metric which describes an homogeneous but anisotropic model and that can be related easily with the flat FLRW metric described in the last section (for a review see for instance \cite{RyanM}).

\begin{figure}[h!]
		\centering
		\includegraphics[width=0.6\textwidth]{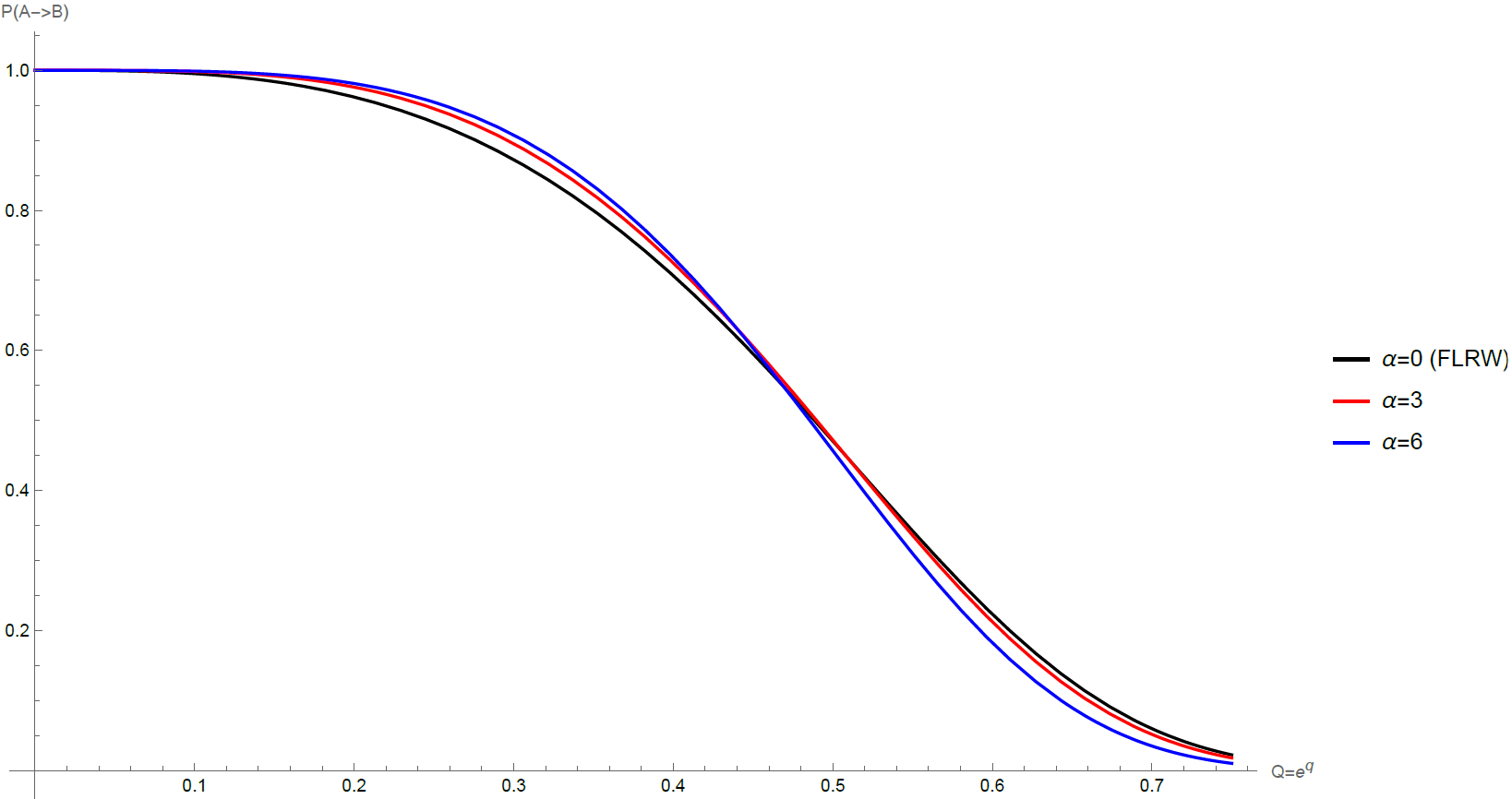}
		\caption{Transition probability for the Bianchi III metric with SUP with $\alpha=6$ (Blue curve), $\alpha=3$ (Red curve) and the isotropy limit with leads to the FLRW flat result when  $\alpha=0$ (Black curve) choosing 2Vol$(X)=1$, $V_{A}=-100$, $V_{B}=-200$ and $T=10$.  }
		\label{F-BianchiCom}
	\end{figure}

The Bianchi III metric can be written in Cartesian coordinates as
	\begin{equation}\label{B3DefMetric}
		ds^2=-N^2(t)dt^2+P^2(t)dx^2+Q^2(t)e^{-2\alpha x}dy^2+U^2(t)dz^2 ,
	\end{equation}
where $\alpha\neq0$ is a constant. The isotropy limit of this metric is
$P(t)=Q(t)=U(t)$ and $\alpha=0$ and results in the flat FLRW metric
(\ref{FLRWFMetric}). The Hamiltonian constraint in this case leads to
	\begin{multline}\label{B3HCAux}
		H= \frac{N}{PQU}\bigg\{\frac{P^2}{4}\pi^2_{P}-\frac{PQ}{2}\pi_{P}\pi_{Q}-\frac{PU}{2}\pi_{P}\pi_{U}+\frac{Q^2}{4}\pi^2_{Q}-\frac{QU}{2}\pi_{Q}\pi_{U}+\frac{U^2}{4}\pi^2_{U}+\frac{1}{2}\pi^2_{\phi}  \\  + Q^2U^2\alpha^2+P^2Q^2U^2V(\phi) \bigg\} \simeq 0 ,
	\end{multline}
where we have factorized the term $(PQU)^{-1}$ for convenience. However as we have done so far it is more convenient to make the change of variables 
	\begin{equation}\label{B3VariablesDef}
		P(t)=e^{p(t)} , \hspace{0.5cm} Q(t)=e^{q(t)} , \hspace{0.5cm} U(t)=e^{u(t)} ,
	\end{equation}
then the Hamiltonian constraint takes the form
	\begin{multline}\label{B3HC}
		H= \frac{N}{e^{p+q+u}}\bigg\{\frac{\pi^2_{p}}{4}+\frac{\pi^2_{q}}{4}+\frac{\pi^2_{u}}{4}-\frac{\pi_{p}\pi_{q}}{2}-\frac{\pi_{p}\pi_{u}}{2}-\frac{\pi_{q}\pi_{u}}{2}+\frac{\pi^2_{\phi}}{2} \\  +  e^{2(q+u)}\left[\alpha^2+e^{2p}V(\phi)\right] \bigg\}\simeq 0 ,
	\end{multline}
where the canonical momenta are defined as
	\begin{equation}\label{B3DefMomenta}
		\pi_{p}=-\frac{\dot{q}+\dot{u}}{N}e^{p+q+u} , \hspace{0.5cm} \pi_{q}=-\frac{\dot{p}+\dot{u}}{N}e^{p+q+u} , \hspace{0.5cm} \pi_{u}=-\frac{\dot{p}+\dot{q}}{N}e^{p+q+u} , \hspace{0.5cm} 
		\pi_{\phi}=\frac{\dot{\phi}}{N}e^{p+q+u} ,
	\end{equation}
and as usual $\pi_{N}=0$ and therefore we only focus on the terms inside brackets. We will consider the WDW equation obtained after doing canonical quantization with variables in minisuperspace that obey a standard uncertainty relation as usual. Following the procedure of \cite{LVTAU} with $s$ chosen as in (\ref{ChooseS}) we obtain
	\begin{multline}\label{B3GammaAux}
		\pm\Gamma=-2{\rm Vol}(X)i\left[\int_{0}^{\bar{s}-\delta s}\frac{f}{C(s)}\bigg\rvert_{\phi=\phi_{B}}ds-\int_{0}^{\bar{s}-\delta s}\frac{f}{C(s)}\bigg\rvert_{\phi=\phi_{A}}ds\right]\\+\frac{{\rm Vol}(X)}{\hbar}e^{2(\bar{p}+\bar{q}+\bar{u})}T ,
	\end{multline}
where the tension term was defined as usual by
	\begin{equation}\label{B3DefTension}
		{\rm Vol}(X)e^{2(\bar{p}+\bar{q}+\bar{u})}T=-2{\rm Vol}(X)i\int_{\bar{s}-\delta s}^{\bar{s}+\delta s}\frac{V(\phi)-V_{A}}{C(s)}e^{2(p+q+u)}ds ,
	\end{equation}
and from (\ref{B3HC}) we see that in this case $f(p,q,u)=e^{2(q+u)}\left[\alpha^2+e^{2p}V(\phi)\right]$. The integrals are performed by solving the general system of equations. In this case the coordinates on minisuperspace turn out to be related as a consequence of the semi-classical approximation. In the region where the scalar field is constant they are related by
	\begin{equation}\label{B3RelationsC}
		q=u=\ln\sqrt{\frac{2\alpha^2}{V}+e^{2p}} ,
	\end{equation}
where the integration constants were absorbed in the coordinates by a redefinition. We note from here that the isotropy limit turns out to be just the limit $\alpha\to0$. With this the integrals on the parameter $s$ can be put into integrals of any of the three coordinates of minisuperspace. Thus solving the system of equations we obtain finally in the thin wall limit
	\begin{multline}\label{B3ProbComp}
		\pm\Gamma=\pm2{\rm Vol}(X)i\left[\sqrt{V_{B}}\int_{-\infty}^{\bar{p}}e^{p}\sqrt{\left(e^{2p}+\frac{\alpha^2}{V_{B}}\right)\left(3e^{2p}+\frac{4\alpha^2}{V_{B}}\right)}dp \right. \\ \left. -\sqrt{V_{A}}\int_{-\infty}^{\bar{p}}e^{p}\sqrt{\left(e^{2p}+\frac{\alpha^2}{V_{A}}\right)\left(3e^{2p}+\frac{4\alpha^2}{V_{A}}\right)}dp\right]+{\rm Vol}(X)e^{2\bar{p}}\left(e^{2\bar{p}}+\frac{2\alpha^2}{V_{B}}\right)^2T ,
	\end{multline}
where we have chosen $p_{0}=-\infty$ and used the thin wall limit to write the tension term in terms of only $\bar{p}$ demanding continuity of $p$. This result is the same as the one obtained in \cite{LVTAU} when the appropriate change of coordinates is made. However in the following it will be more useful to express this result in terms of the variable $q$ which leads to
	\begin{multline}\label{B3Prob}
		\pm\Gamma=\pm2{\rm Vol}(X)i\left[\sqrt{V_{B}}\int_{q_{0}}^{\bar{q}}e^{2q}\sqrt{\frac{\left(e^{2q}-\frac{\alpha^2}{V_{B}}\right)\left(3e^{2q}-\frac{2\alpha^2}{V_{B}}\right)}{e^{2q}-\frac{2\alpha^2}{V_{B}}}}dq \right. \\ \left. -\sqrt{V_{A}}\int_{q_{0}}^{\bar{q}}e^{2q}\sqrt{\frac{\left(e^{2q}-\frac{\alpha^2}{V_{A}}\right)\left(3e^{2q}-\frac{2\alpha^2}{V_{A}}\right)}{e^{2q}-\frac{2\alpha^2}{V_{A}}}}dq\right]+{\rm Vol}(X)e^{4\bar{q}}\left(e^{2\bar{q}}-\frac{2\alpha^2}{V_{B}}\right)T ,
	\end{multline}
where by consistency with the relations (\ref{B3RelationsC}) we have that for positive values of the potential minima $q_{0}$ cannot be chosen to be $-\infty$ and therefore since this expression is valid in the range $\bar{q}\geq q_{0}$ we can no longer have access to the UV limit $q_{0}\to-\infty$, it is only when $\alpha\to0$ that we can recover the access to this area. In the limit $\alpha\to0$ this result reduces to (\ref{FLRWFProb}) as expected.  Thus we note that for positive values of the potential the expression that is valid for all range of values including the UV limit and therefore it is more suitable to plot and study the behaviour and effect of anisotropy is (\ref{B3ProbComp}). In this case the first term does not contribute and it is found that as the anisotropy constant squared $\alpha^2$ increases, the probability decreases and falls faster to $0$ \cite{LVTAU}. In the other hand, for negative values of the potential minima, the relations (\ref{B3RelationsC}) tell us that $e^{p_{0}}$ can no longer be chosen to be zero, but choosing $q_{0}=-\infty$ is possible, therefore in this case the expression in terms of $q$ (\ref{B3Prob}) is well defined in all the range including the UV limit and thus it is more suitable to study the behaviour of the probability with this variable. We note from the tension term that as $\alpha^2$ increases, this term will increase as well similarly to what happened in (\ref{B3ProbComp}). However when the potential minima are negative the first term will contribute and therefore the behaviour will not be due to the tension term only. In order to visualize what is the effect of the anisotropy on the probability when the potential minima are negative we plot the transition probability coming from (\ref{B3Prob}) for different values of $\alpha$ and choosing $2{\rm Vol}(X)=1$, $V_{A}=-100$, $V_{B}=-200$ and $T=10$. In Figure \ref{F-BianchiCom} we show this behaviour. If we choose the plus sign in the right hand side of (\ref{B3Prob}) we can obtain the same behaviour as the one encountered with the variable $p$ that is, as $\alpha$ increases the probability decreases, however there will be a region near $e^{q}=0$ where the flat limit $\alpha\to0$ would not have a proper behaviour since it would lead to values greater than $1$. Therefore in order to obtain well defined probabilities we choose the minus sign. That leads us to a scenario where the tension term has the same behaviour just described but the first term has the opposite behaviour and then as shown in the figure we obtain two regions with opposite behaviour. Initially when the first term is more important the probability  increases with $\alpha$ but as $e^{q}$ keeps growing the tension term grows and eventually it becomes dominant and we recover the same behaviour as we had with the variable $p$. 

\subsection{Considering a GUP and anisotropy}
Now let's consider a set of coordinates denoted with a subscript $g$ that obey the GUP (\ref{RelationGUP}) and use for the momenta the corresponding representation as in Eq. (\ref{MomOperatorsGUP}). We note that in this case
	\begin{equation}\label{B3P2}
		\mathcal{P}^2=\frac{\pi^2_{p}}{4}+\frac{\pi^2_{q}}{4}+\frac{\pi^2_{u}}{4}-\frac{\pi_{p}\pi_{q}}{2}-\frac{\pi_{p}\pi_{u}}{2}-\frac{\pi_{q}\pi_{u}}{2}+\frac{\pi^2_{\phi}}{2} .
	\end{equation}	
Thus using the relations between both sets of variables (\ref{RelationsCoordinates}) and the Zassenhaus formula (\ref{ZSFormula})  we obtain in this case
	\begin{equation}\label{B3Expansion1}
		e^{2(q_{g}+u_{g})}\simeq e^{2(q+g)}\left[1+2i\gamma^2(q+u)\pi_{p}+2\gamma^2(q+u)\mathcal{P}^2\right],
	\end{equation}
	\begin{multline}\label{B3Expansion2}
		e^{2(p_{g}+q_{g}+u_{g})}V(\phi_{g})\simeq e^{2(1+\gamma^2)(p+q+u)}V(\phi)\left[1+i\gamma^2(p+q+u)(\pi_{p}+\pi_{q}+\pi_{u})\right. \\ \left. +2\gamma^2(p+q+u)\mathcal{P}^2\right] .
	\end{multline}
Then the resulting WDW equation takes the form
	\begin{multline}\label{B3GUPHC}
		\left\{H[p,q,u,V(\phi)]\left[\frac{\pi^2_{p}}{4}+\frac{\pi^2_{q}}{4}+\frac{\pi^2_{u}}{4}-\frac{\pi_{p}\pi_{q}}{2} -\frac{\pi_{p}\pi_{u}}{2}-\frac{\pi_{q}\pi_{u}}{2}+\frac{\pi^2_{\phi}}{2}\right]\right. \\ \left. +L_{1}[p,q,u,V(\phi)](\pi_{q}+\pi_{u}) +\left(L_{1}[p,q,u,V(\phi)]+L_{2}[p,q,u]\right)\pi_{p} \right. \\ \left. +\alpha^2e^{2(q+u)}+e^{2(1+\gamma^2)(p+q+u)}V(\phi)\right\}\Psi(p,q,u,\phi)\simeq 0,
	\end{multline}
where we have defined
	\begin{equation}\label{B3GUPHDef}
		H[p,q,u,V(\phi)]=1+2\gamma^2\alpha^2e^{2(q+u)}(q+u)+2\gamma^2e^{2(1+\gamma^2)(p+q+u)}(p+q+u)V(\phi) ,
	\end{equation}
	\begin{equation}\label{B3GUPL1Def}
		L_{1}[p,q,u,V(\phi)]=i\gamma^2e^{2(1+\gamma^2)(p+q+u)}(p+q+u)V(\phi) ,
	\end{equation}
	\begin{equation}\label{B3GUPL2Def}
		L_{2}[p,q,u]=2i\gamma^2\alpha^2e^{2(q+u)}(q+u) .
	\end{equation}
Comparing with the general form of the Hamiltonian constraint in (\ref{GeneralHamiltonian}) we note that in this case
	\begin{equation}\label{B3GUPMetricD}
		\{G^ {MN}\}= \frac{H[p,q,u,V(\phi)]}{2} \left( \begin{array}{cccc}
			1 & -1 & -1 & 0 \\
			-1 & 1 & -1 & 0 \\
			-1 & -1 & 1 & 0  \\
			0 & 0  & 0 & 2   \end{array} \right) ,
	\end{equation}
	\begin{equation}\label{B3GUPWDef}
		\{W^{M}\}=L_{1}[p,q,u,V(\phi)](1,1,1,0)+L_{2}[p,q,u](1,0,0,0) ,
	\end{equation}
	\begin{equation}\label{B3GUPfDef}
		f(p,q,u,\phi)=\alpha^2e^{2(q+u)}+e^{2(1+\gamma^2)(p+q+u)}V(\phi) .
	\end{equation}
Choosing $s$ as in (\ref{ChooseS}) and following the same procedure as in the previous sections we obtain
	\begin{multline}\label{B3GUPGamma}
		\pm\Gamma=-2{\rm Vol}(X)i\left[\int_{0}^{\bar{s}-\delta s} \left(\frac{f}{C(s)}+W_{M}\frac{d\Phi^M}{ds}\right)\bigg\rvert_{\phi=\phi_{B}}ds\right. \\ \left. -\int_{0}^{\bar{s}-\delta s}\left(\frac{f}{C(s)}+W_{M}\frac{d\Phi^M}{ds}\right)\bigg\rvert_{\phi=\phi_{A}}ds \right]+\frac{{\rm Vol}(X)}{\hbar}e^{2(1+\gamma^2)(\bar{p}+\bar{q}+\bar{u})}T ,
	\end{multline}
where we have defined as usual the tension term
	\begin{equation}\label{B3GUPTensionDeg}
		{\rm Vol}(X)e^{2(1+\gamma^2)(\bar{p}+\bar{q}+\bar{u})}T=-2{\rm Vol}(X)i\int_{\bar{s}-\delta s}^{\bar{s}+\delta s}\frac{V(\phi)-V_{A}}{C(s)}e^{2(1+\gamma^2)(p+q+u)}ds .
	\end{equation}
The integrals in (\ref{B3GUPGamma}) are obtained after solving the system of equations (\ref{System1}) and (\ref{System2}) in the case in which the scalar field is constant. In this case such system is written as
	\begin{multline}\label{B3Systempo}
		\frac{dp}{ds}=\frac{2{\rm Vol}(X)}{C^2(s)}H(p,q,u,V)\left[(1+\gamma^2)e^{2(1+\gamma^2)(p+q+u)}V+2\alpha^2e^{2(q+u)}\right] \\ -\frac{{\rm Vol}(X)}{C(s)}\left[\frac{dp}{ds}+\frac{1}{2}\frac{dq}{ds}+\frac{1}{2}\frac{du}{ds}\right]\frac{8i\gamma^4\alpha^2e^{2\left[q+u+(1+\gamma^2)(p+q+u)\right]}V[2\gamma^2(p+q+u)(q+u)-p]}{H[p,q,u,V]} ,
	\end{multline}
	\begin{multline}\label{B3Systemqo}
		\frac{dq}{ds}=\frac{2{\rm Vol}(X)}{C^2(s)}H[p,q,u,V](1+\gamma^2)e^{2(1+\gamma^2)(p+q+u)}V \\ +\frac{2{\rm Vol}(X)}{C(s)}\left[\frac{dq}{ds}+\frac{du}{ds}\right]\frac{4i\gamma^4\alpha^2e^{2\left[q+u+(1+\gamma^2)(p+q+u)\right]}V[2\gamma^2(p+q+u)(q+u)-p]}{H[p,q,u,V]} ,
	\end{multline}
	\begin{multline}\label{B3Systemuo}
		\frac{du}{ds}=\frac{2{\rm Vol}(X)}{C^2(s)}H[p,q,u,V](1+\gamma^2)e^{2(1+\gamma^2)(p+q+u)}V \\ +\frac{2{\rm Vol}(X)}{C(s)}\left[\frac{dq}{ds}+\frac{du}{ds}\right]\frac{4i\gamma^4\alpha^2e^{2\left[q+u+(1+\gamma^2)(p+q+u)\right]}V[2\gamma^2(p+q+u)(q+u)-p]}{H[p,q,u,V]} ,
	\end{multline}
where $V$ represents $V_{A}$ or $V_{B}$ depending in which path is being considered. However we note that in the three equations the second term has an overall dependence on $\gamma^4$ which is smaller than the first factor by two powers of $\gamma$. Therefore in the three cases we can ignore the second term and thus the system of equation simplifies to
		\begin{equation}\label{B3Systemp}
			\frac{dp}{ds}\simeq\frac{2{\rm Vol}(X)}{C^2(s)}H[p,q,u,V]\left[(1+\gamma^2)e^{2(1+\gamma^2)(p+q+u)}V+2\alpha^2e^{2(q+u)}\right]  ,
		\end{equation}
		\begin{equation}\label{B3Systemq}
			\frac{dq}{ds}\simeq\frac{2{\rm Vol}(X)}{C^2(s)}H[p,q,u,V](1+\gamma^2)e^{2(1+\gamma^2)(p+q+u)}V ,
		\end{equation}
		\begin{equation}\label{B3Systemu}
			\frac{du}{ds}\simeq\frac{2{\rm Vol}(X)}{C^2(s)}H[p,q,u,V](1+\gamma^2)e^{2(1+\gamma^2)(p+q+u)}V  .
		\end{equation}
From this we note that once again the variables are related to each other, in this case we obtain
	\begin{multline}\label{B3GUPRelationsC}
		\frac{1}{(1+\gamma^2)e^{2(1+\gamma^2)(p+q+u)}V+2\alpha^2e^{2(q+u)}}\frac{dp}{ds}=\frac{1}{(1+\gamma^2)e^{2(1+\gamma^2)(p+q+u)}V}\frac{dq}{ds}\\=\frac{1}{(1+\gamma^2)e^{2(1+\gamma^2)(p+q+u)}V}\frac{du}{ds} ,
	\end{multline}
which can be solved to give the following relations
	\begin{equation}\label{B3GUPRelationsF}
		u=q , \hspace{0.5cm} p=\ln\left[\left(e^{2(1+\gamma^2)q}-\frac{2\alpha^2}{V(1+3\gamma^2)}e^{-4\gamma^2q}\right)^{\frac{1}{2(1+\gamma^2)}}\right] ,
	\end{equation}
where we have ignored the integration constants since they can be absorbed by a redefinition of the coordinates. Thus we obtain from (\ref{System1}) a quadratic equation for $C(s)$ with solution
	\begin{equation}\label{B3GUPSolCA}
		C(s)=\frac{M_{1}}{2f}\pm\frac{\sqrt{M_{1}^2+4M_{2}f}}{2f} ,
	\end{equation}
where
	\begin{multline}\label{B3GUPDefM1}
		M_{1}=4{\rm Vol}(X)i\gamma^2e^{2(1+\gamma^2)(p+q+u)}V\left[3(1+\gamma^2)e^{2(1+\gamma^2)(p+q+u)}(p+q+u)V\right. \\ \left. +2\alpha^2e^{2(q+u)}\left[p+q+u+(1+\gamma^2)(q+u)\right]\right] ,
	\end{multline}
	\begin{multline}\label{B3GUPDefM2}
		M_{2}=4{\rm Vol}^2(X)H[p,q,u,V](1+\gamma^2)e^{2(1+\gamma^2)(p+q+u)}V\left[3(1+\gamma^2)e^{2(1+\gamma^2)(p+q+u)}V\right. \\ \left. +4\alpha^2e^{2(q+u)}\right].
	\end{multline}
From this we note that $M_{1}^2$ has a global factor of $\gamma^4$ and therefore it can be ignored with respect to $4M_{2}f$. Thus we obtain
	\begin{equation}\label{B3GUPSolC}
		C(s)\simeq\frac{M_{1}}{2f}\pm\sqrt{\frac{M_{2}}{f}} ,
	\end{equation}
and thus when the scalar field is constant we obtain up to second order in $\gamma$ the following equation
	\begin{equation}\label{B3GupAction}
		\int \left(\frac{f}{C(s)}+W_{M}\frac{d\Phi^M}{ds}\right)ds\simeq\pm\int \sqrt{F_{III}(q,V)}dq -\frac{i\gamma^2}{1+\gamma^{2}}\int G_{III}(q,V)dq , 
	\end{equation}
where we have defined the functions	
	\begin{equation}\label{B3GUPDefF}
		F_{III}(q,V)=\frac{\left[3(1+\gamma^2)Ve^{6(1+\gamma^2)q}-\frac{2\alpha^2(1-3\gamma^2)}{1+3\gamma^2}e^{4q}\right]\left[Ve^{6(1+\gamma^2)q}-\frac{1-3\gamma^2}{1+3\gamma^2}\alpha^2e^{4q}\right]}{H[q,V]\left[V(1+\gamma^2)e^{6(1+\gamma^2)q}-\frac{2\alpha^2(1+\gamma^2)}{1+3\gamma^2}e^{4q}\right]} ,
	\end{equation}
	\begin{equation}\label{B3GUPDefG}
		G_{III}(q,V)=\frac{3V(1+\gamma^2)(2q+p(q))e^{6(1+\gamma^2)q}+\frac{4\alpha^2}{1+3\gamma^2}\left[(3\gamma^4+4\gamma^2-1)q-p(q)\right]e^{4q}}{H[q,V]} .
	\end{equation}
In the above equations $p(q)$ was written in such a way to remind us that $p$ is related to $q$ as stated in the relations (\ref{B3GUPRelationsF}) and $H[q,V]$ means the function defined in (\ref{B3GUPHDef}) with the relations  taken into account which leads to
	\begin{equation}\label{B3HFunctionF}
		H[q,V]=1+2\gamma^2V(2q+p(q))e^{6(1+\gamma^2)q}-\frac{4\gamma^2\alpha^2}{1+3\gamma^2}\left[(1-3\gamma^2)q+p(q)\right]e^{4q} .
	\end{equation}

In this way we can also reduce the tension term in (\ref{B3GUPGamma}) in the thin wall limit since in that case those relations are valid for all values of $s$ with its corresponding integration constants and we can ask for continuity in $q$. In this way the tension term will also be written in terms of just $\bar{q}$ and the potential minima. Thus we obtain finally in the thin wall limit
	\begin{multline}\label{B3GUPProb}
		\pm\Gamma\simeq\pm2{\rm Vol}(X)i\left[\int_{q_{0}}^{\bar{q}}\sqrt{F_{III}(q,V_{B})}dq-\int_{q_{0}}^{\bar{b}}\sqrt{F_{III}(q,V_{A})}dq\right]\\-\frac{2{\rm Vol}(X)\gamma^2}{(1+\gamma^2)}\left[\int_{q_{0}}^{\bar{q}}G_{III}(q,V_{B})dq-\int_{q_{0}}^{\bar{q}}G_{III}(q,V_{A})db\right]\\+{\rm Vol}(X)\left[e^{6(1+\gamma^2)\bar{q}}-\frac{2\alpha^2}{V_{B}(1+3\gamma^2)}e^{4\bar{q}}\right]T ,
	\end{multline}
where in each integral $p$ must be related to $q$ as in (\ref{B3GUPRelationsF}) with the corresponding value of the potential since as each integral corresponds to a different path in both cases the integration constants can be ignored. We note that as it happened in the case without a GUP the value of $q_{0}$ can only be chosen to be $-\infty$ when the potential minima are negative, and since in this case it is difficult to express the integral in terms of the variable $p$ we will only choose negative potential minima in order to compare this result to the one obtained without a GUP. We also note that once again the transition probability is described by only two parameters, in this case $\bar{q}$ and $T$. The first term could be ignored if the integrals turn out to be real since only the real part of this expression will contribute to the transition probability whereas the second part will always have to be considered since they are always real.

We can see that in the limit $\gamma\to0$ this expression reduces to Eq. (\ref{B3Prob}) as expected and in the isotropy limit $\alpha\to0$ this reduces to the result found for the flat FLRW metric (\ref{FLRWGUPFProb}) as expected as well. Thus this result fulfils the two limits that we expected and give us a transition probability that has to be evaluated numerically but depends on the same number of parameters as before.

In order to obtain well defined probabilities we choose in (\ref{B3GUPProb}) the minus sign on the right hand side and the plus sign on the left. When we change the parameter $\gamma$ the behaviour found is in agreement whit the earlier result for the FLRW metric that is the effect of considering a GUP is that the probability is increased at first but then it decreases faster and eventually the roles are interchanged and the probability decreases with $\gamma$. On the other hand when we vary the anisotropy parameter $\alpha$ the behaviour found is the same as the one described by Figure \ref{F-BianchiCom} when it was assumed a SUP, that is we obtain the same two regions as explained before. It is interesting to note that varying both parameters of GUP $\gamma$ and anisotropy $\alpha$ leads to a very similar behaviour although these parameters are independent to each other. In any case, all the probabilities have the same limits that the one encountered without a GUP and with isotropy, that is they all go to $1$ when $q\to-\infty$ and as $Q$ increases they decrease going to zero.

\section{Some other possible extensions}
\label{GUP-EUP}
In the present article we have considered the GUP and its application to the computation of transition probabilities in the early universe in the context of the WDW equation. It is based on a modification of the Heisenberg (Standard) Uncertainty Principle (SUP) of the form 
\begin{equation}
\Delta q_i \cdot \Delta p_j \geq  {\hbar \over 2}\delta_{ij}\big[1 + \gamma^2(\Delta p)^2\big],
\end{equation}
where $\gamma$ is the GUP parameter and we consider it on the variables of minisuperspace.  The associated commutator algebra is given by 
\begin{equation}
[q_i,p_j] = i \hbar\delta_{ij}\big[1 + \gamma^2 p^\ell p_\ell \big],
\end{equation}
where a sum over $\ell$ is understood. In the {\it momentum representation} we have
\begin{equation}
q_i = i\hbar(1+\gamma^2 p^\ell p_\ell){\partial \over \partial p_i} = (1+\gamma^2 p^\ell p_\ell)q'_i.
\label{qes}
\end{equation}
Here  ${q'}_i  =i\hbar {\partial \over \partial p_i}$ and  thus the commutator algebra is given by  $[q'_i, p_j] = i \hbar \delta_{ij}$. In this representation it is known that: $[p_{i},p_{j}]=0$ and $[q_i,q_j] \not= 0$ and the later is explicitly given by
\begin{equation}
[q_i,q_j] = 2 i \hbar \gamma^2(p_iq_j -p_jq_i).
\end{equation}
Thus a noncommutativity of the minisuperspace is implicit in the description. 

Then working in the momentum representation the WDW equation is written as follows 
\begin{equation}
H\bigg(p_i,i\hbar(1+\gamma^2 p^\ell p_\ell){\partial \over \partial p_i}\bigg) \phi(p)= 0.
\end{equation}
In the potential part of the hamiltonian $H$, $V=V(q)$ where $q$ is given through Eq. (\ref{qes}) in terms of $q'$. Taylor expanding $V(q)$, keeping terms up to second order in $\gamma$ and momenta and taking a coordinate representation of the WDW equation for variables $p$ and $q'$ leads to a second order differential equation which can be solved exactly under certain considerations. This procedure was applied in Refs.  \cite{GUP,GUPHL} to some specific minisuperpace models of the WDW equation. Doing the same procedure in the present article we studied the semi-classical transition probabilities for different models in an early universe.  

Now we briefly describe how another possible extension of the uncertainty principle (called the Extended Uncertainty Principle (EUP)) can be incorporated into the quantum cosmological context of the present article. It is known that quantum mechanics in an (anti)-de Sitter background can modify the uncertainty principle by a term which depends on a cosmological constant $\Lambda$ \cite{Park:2007az,Mignemi:2009ji,Gine:2020izd} 
\begin{equation}
\Delta q_i \cdot \Delta p_j \geq   {\hbar\over 2} \delta_{ij}\big[1 + \lambda^2(\Delta q)^2\big],
\end{equation}
where $\lambda^2 =\ell_H^{-2} = -\Lambda/3$ is the EUP parameter. In this case the commutator algebra reads
\begin{equation}
[q_i,p_j] = i \hbar \delta_{ij}\big[1 + \lambda^2 q^\ell q_\ell \big].
\end{equation}
In the dual {\it coordinate representation} we have  
\begin{equation}
p_i = i\hbar(1+\lambda^2 q_\ell q_\ell){\partial \over \partial q_i} = (1+\lambda^2 q^\ell q_\ell)p'_i,
\label{pes}
\end{equation}
where $p'_i = -i \hbar {\partial \over \partial q_i}$  and $q_i$ fulfils the commutator algebra $[q_i, p'_j] = i \hbar \delta_{ij}$. It is easy to see that in this representation: $[q_{i},q_{j}]=0$ while momenta do not commute $[p_{i},p_{j}]\not= 0$ i.e.
\begin{equation}
[p_i,p_j] = 2 i \hbar \lambda^2(q_jp_i -q_ip_j).
\end{equation}
Consequently we are in the dual coordinate representation. The implementation in the WDW equation yields
\begin{equation}
H\bigg(-i\hbar(1+\lambda^2 q^\ell q_\ell){\partial \over \partial q_i},q_i\bigg) \psi(q)= 0.
\end{equation}
Thus in this representation the potential $V=V(q)$ will remain without change but the momentum $p$ will obey transformation (\ref{pes}). Writing the WDW equation in terms of momenta $p'$, considering the momentum representation for variables $p'$ and $q$ and keeping terms at most order $\lambda^2$ one would obtain a second order differential equation that would still contain interesting information of the relevant system with a EUP. It would be interesting to consider some models in quantum cosmology in the momentum representation and study transition probabilities in this context. It is worth mentioning that the implementation of the EUP in variables of minisuperspace has a different meaning that the original interpretation of quantum mechanics in (anti)-de Sitter space \cite{Park:2007az,Mignemi:2009ji,Gine:2020izd}. However the relation between both descriptions is a very interesting subject to be explored.

Finally it is worth to consider the case of the combined effects of GUP and EUP. In this situation we have 
\begin{equation}
\Delta Q_i \cdot \Delta P_j \geq  \hbar \delta_{ij}\big[1 +  \gamma^2(\Delta P)^2 + \lambda^2(\Delta Q)^2\big],
\end{equation}
where we have the combined effects of the parameters $\hbar$, $\gamma^2$ and $\lambda^2$.
The associated commutator algebra is given by
\begin{equation}
[Q_i,P_j] = i \hbar \delta_{ij}\big[1 + \gamma^2 P^r P_r + \lambda^2 Q^s Q_s \big].
\end{equation}
In this case it would be interesting to find a double transformation for $Q_i$ and $P_i$ (in terms of $Q'_i$ and $P'_i$), such that $[Q'_i,P'_j] = i \hbar \delta_{ij}$ and additionally $[Q_i,Q_j]\not=0$ and $[P_i,P_j]\not=0$. One possible candidate to carry out such description would be the {\it phase space representation} \cite{GoTorres1,GoTorres2,Curtright:2011vw} in where the WDW equation have the form
\begin{equation}
H \bigg(-i\hbar(1+\lambda^2 Q^s Q_s){\partial \over \partial Q_i},i\hbar(1+\gamma^2 P^r P_r){\partial \over \partial P_i}\bigg) \Psi(P,Q)= 0.
\end{equation}
It would be interesting to see if this proposal is viable and in such case apply it to the description of quantum cosmological models using the WDW equation in phase space representation and compute the corresponding transition probabilities. 

\section{Final Remarks}
\label{S-FinalRemarks}
In the present article we implemented a general procedure to compute the vacuum transition probabilities between two minima of a scalar field potential by solving a general WDW equation and using a WKB approach. The general form of the WDW equation considered  contain terms of linear momenta that represents a generalization of the method introduced in \cite{Cespedes:2020xpn,LVTAU} and it was found to be useful in the case of the WDW equation that appeared when we considered a GUP in the coordinates of minisuperspace including the scalar field. We studied these transitions for an homogeneous and isotropic FLRW metric with a positive curvature as well as with null curvature, and we also studied the scenario of the Bianchi III metric. In all cases analytic expressions were found for the transition probabilities after making an approximation up to second order in momenta and in the GUP parameter $\gamma$.

We first studied the transition probabilities for the FLRW metrics with a SUP. In order to simplify the GUP analysis a change of variables was carried out and an appropriate factorization was considered. In the FLRW flat case the result is in agreement with the result found in \cite{LVTAU}. However since the factorization used in this work for the closed FLRW metric is different from the one used in the work just mentioned the result is slightly different but the general behaviour is not changed. Then a GUP was taken into account using the procedure described in \cite{GUPHL} and by using the general method we obtained the transition probability in this case as well. The probability was also described in terms of just two parameters in both cases, namely, the scale factor $\bar{a}$ and the tension $T$. However for the closed FLRW metric the extremizing procedure was found to be troublesome and the expression involved in the probability was written in terms of integrals that have to be evaluated numerically. The functions inside the integrals however have singular points for some value of the scale factor depending on the choice of the potential minima and therefore in order to have well defined probabilities we had to avoid those points. It was found in both cases that the effect of considering a GUP does not change the general behaviour, that is we still have access to the UV limit $\bar{a}\to0$ and obtain that the probability goes to $1$ and then it decreases. However two regions are found, when the scale factor is small enough as the GUP parameter $\gamma$ increases so does the probability but these curves decrease faster and when the scale factor is big enough the behaviour is flipped, that is in an infrared region as $\gamma$ increases the probability decreases. For the FLRW closed case we showed the plot of the probabilities in the UV region in Figure \ref{F-FLRW} and  in the region where the change of behaviour is encountered in Figure \ref{F-FLRWExtra}. 

For the Bianchi III metric we first studied once again the transition probabilities when the SUP is considered and used a useful change of coordinates. The results found are in complete agreement with the results of \cite{Mansouri,LVTAU}. When the potential minima are positive the probability was found to be described correctly with the variable $p$ and the effect of anisotropy was found to be that as the anisotropy constant $\alpha^2$ increases the probability decreases as it was known. However, in the case in which the potential minima are negative the probability is best described in terms of another of the variables and in that case the effect of anisotropy was not so simple. In this scenario two regimes appeared, when the corresponding scale factor $Q=e^{q}$ is small enough the probability increases with $\alpha^2$ but it decreases faster and when this variable is big enough the effect of anisotropy found earlier is recovered. Let us point out that this change of the effect of anisotropy is not due to the change of variables, the same scenario is present in the expressions of \cite{LVTAU} and it comes from the fact that even though the coordinates of minisuperspace are related due to the semi-classical approximation used, there are still two variables to express the transition probability and thus two possible behaviours. Then we considered a GUP as in the previous case. The coordinates of minisuperspace were also found to be related but in a more complicated way that involves both $\alpha$ and $\gamma$ and the transition probability was found to be described in terms of just two parameters as well. However in this case it is only possible to express the probability analytically in terms of the variable $q$ and therefore we only used negative potential minima. The expression found has to be evaluated numerically but it fulfils the two expected limits. That is in the limit $\alpha\to 0$ it reduces to the flat FLRW result and in the limit $\gamma\to0$ it reduces to the Bianchi III with SUP result. The behaviour found is consistent with the previous results when the parameters $\alpha$ and $\gamma$ are varied. That is in both cases there are two regions, first the probability increases  then it decreases with the varying parameter. We point out that although both behaviours are equivalent, they come from different functions and arguments. 

In all metrics considered here it was found that the effect of considering a  GUP is that the probability increases at first but it decays faster and then it decreases with the parameter $\gamma$. Thus in the presence of a GUP the transitions are expected to happen when the scale factor is in general smaller than in the case with a SUP. Although the point of maximum probability in both cases is the singularity $\bar{a}\to0$ ($\bar{Q}\to0$ in Bianchi III), the GUP scenario at first decays slower and thus for values of the scale factor different from zero but very small, that is in the UV region, the probability coming from the GUP is bigger than in the other case. Therefore taking the interpretation proposed in \cite{Cespedes:2020xpn} and referred to in \cite{LVTHL} that these expressions give us probabilities of creating universes by vacuum transitions, the probability of creating a universe with a non-zero small size is enhanced when a GUP is considered compared to the SUP case which indicates that the idea of a minimal measurable length derived from the GUP is present in the transition probability even if the GUP is taken in the minisuperspace with the gravitational degrees of freedom as well as the scalar field. Although we have to point out that contrary to what happened in \cite{LVTHL} the maximum probability in this case is still obtained in the singularity point $\bar{a}\to0$ ($\bar{Q}\to0$) and therefore there is not a full resolution of the singularity at least at the semi-classical level. It would be interesting to take more terms in the WKB expansion which will presumably lead us to quantum corrections of the results found and look for additional changes when the GUP is considered which could modify in particular the UV region.

In Sec. \ref{GUP-EUP} we give a very short overview of the GUP approach considered in this work. We also describe additional variations of the commutation relations used in the literature. In particular the Extended Uncertainty Principle (EUP) which leads to a minimum measurable value for the momentum, has been shown to arise naturally when the quantization procedure is performed in an (anti)-de Sitter background (see for instance,  \cite{Park:2007az,Mignemi:2009ji,Gine:2020izd} and references therein). It may be of interest to consider the transition probabilities in this scenario as well. In this case the way in which the variables can be related (\ref{RelationsCoordinates}) will be modified and thus the resulting WDW equation will be different as we briefly mentioned in Section \ref{GUP-EUP}.  However if we also keep terms up to second order on the momenta in the resulting equation, the general form that the WDW equation would have of the same form considered here in (\ref{WDWEqGeneral}) and thus the general method presented in Sec. \ref{S-GeneralMethod} should be applicable in this scenario as well. We have to point out however that in the context of the present article the EUP should be taken in the coordinates and momenta of the minisuperspace and thus it does not represent the same scenario of the works just mentioned \cite{Park:2007az,Mignemi:2009ji,Gine:2020izd}.

 \vspace{1cm}
\centerline{\bf Acknowledgments} \vspace{.5cm} D. Mata-Pacheco would
like to thank CONACyT for a grant.



\end{document}